\documentclass[preprint,12pt,authoryear]{elsarticle}

\usepackage{float}      
\usepackage{graphicx}   
\usepackage{pdflscape}  
\usepackage{afterpage}  
\usepackage{amssymb}    
\usepackage{amsthm}     
\usepackage{amsmath}    
\usepackage{lineno}     
\usepackage{booktabs}   
\usepackage{rotating}   
\usepackage{subfigure}  
\usepackage{threeparttable}             
\usepackage[margin=0.85in]{geometry}    
\usepackage[skip=0.6\baselineskip]{caption} 
\usepackage{array}      
\newcolumntype{L}[1]{>{\raggedright\let\newline\\\arraybackslash\hspace{0pt}}m{#1}}
\newcolumntype{C}[1]{>{\centering\let\newline\\\arraybackslash\hspace{0pt}}m{#1}}
\usepackage[dvipsnames]{xcolor}
\usepackage{hyperref}                   
\hypersetup{colorlinks=true}

\journal{\textit{Reliability Engineering \& System Safety}. Please cite: \href{https://doi.org/10.1016/j.ress.2026.113194}{10.1016/j.ress.2026.113194}}

\begin{document}

\begin{frontmatter}

\title{Antifragile perimeter control: Thriving on disruptions through reinforcement learning}

\author[a]{Linghang Sun} 
\author[a]{Michail A. Makridis\corref{cor1}} \ead{michail.makridis@ivt.baug.ethz.ch}
\author[a]{Alexander Genser}
\author[b]{Cristian Axenie}
\author[c]{Margherita Grossi}
\author[a]{Anastasios Kouvelas}

\affiliation[a]{organization={Institute for Transport Planning and Systems, ETH Zürich},
            city={Zurich},
            postcode={8093}, 
            country={Switzerland}}
\affiliation[b]{organization={Computer Science Department and Center for Artificial Intelligence, Technische Hochschule Nürnberg},
            city={Nürnberg},
            postcode={90489}, 
            country={Germany}}
\affiliation[c]{organization={Intelligent Cloud Technologies Lab, Huawei Munich Research Center},
            city={Munich},
            postcode={80992}, 
            country={Germany}}

\cortext[cor1]{Corresponding author.}

\begin{abstract}
The optimal operation of transportation systems is often susceptible to unexpected disruptions. Many established control strategies reliant on mathematical models can struggle with real-world disruptions, leading to significant divergence from their anticipated efficiency. This study integrates the cutting-edge concept of antifragility with learning-based traffic control strategies to optimize urban road network operations under disruptions. Antifragile systems not only withstand and recover from stressors but also thrive and enhance performance in the presence of such adverse events. We develop a deep reinforcement learning framework that incorporates antifragile modules composed of traffic state derivatives and redundancy. Subsequently, it is evaluated in a cordon-shaped transportation network and a case study with real-world data. Promising results highlight that the proposed algorithm provides: (i) superior performance achieving up to $27.6\%$ and $41.9\%$ performance gain over baselines under increasing demand and supply disruptions, (ii) lower distribution skewness under disruptions, demonstrating its relative antifragility against baselines, (iii) effectiveness under limited observability due to real-world data availability constraints, and (iv) the robustness and transferability to be incorporated into recent state-of-the-art RL architectures. The proposed antifragile methodology is generalizable and holds potential for applications beyond traffic engineering, offering integration into control systems exposed to disruptions across various disciplines.
\end{abstract}

 

\end{frontmatter}


\section{Introduction}
\label{sec: introduciton}

Transportation networks serve as vital channels for the movement of people and the flow of goods, and the control of intelligent transportation systems has led to numerous research endeavors and practical implementations. Given that various sorts of disruptions, such as traffic accidents, social events, and adverse weather conditions, often occur unexpectedly in real-world networks, examining the robustness, resilience, reliability, and safety of transportation systems is highly crucial \citep{ganin_resilience_2019, zayandehroodi_improving_2025}. What makes it even more challenging is the continuous growth of motorized traffic. For instance, an approximate 50\% increase in traffic demand can be observed in both the U.S. \citep{us_department_of_transportation_national_2019} and Switzerland \citep{BFS} over the past few decades. Such growth can lead to not only an escalation of congestion but also more traffic accidents \citep{albalate_relationship_2021}. Moreover, climate change has led to frequent and intensified natural hazards \citep{tong_resilience_2026}. To picture the consequences, recent work in \cite{sun_fragile_2024} demonstrates the fragile nature of road transportation networks through mathematical proof, indicating that performance degrades exponentially with linearly increasing magnitudes of disruptions. Therefore, modern transportation systems must be designed to secure a decent level of service even when faced with disruptions of unforeseen magnitudes and their compounding negative effects.

To address such issues, the concept of antifragility has shed light on a promising solution. Echoing the sentiment of the well-known maxim from Nietzsche \textit{What does not kill me makes me stronger}, antifragility was first introduced in the reputable general-audience publication \textit{Antifragile: Things That Gain from Disorder} \citep{taleb2012antifragile}. It has then been mathematically formulated as an academic concept in \cite{taleb_mathematical_2013}, providing insights into designing systems that can benefit from disruptions and perform better in growing volatility and randomness. Its counterpart concept, \textit{fragility}, can be traced back to complexity theory \citep{vespignani_fragility_2010}, indicating a cascading effect of interdependent variables in complex networks, such as in transportation networks \citep{cats_quantifying_2021}. Ever since the concept was proposed, antifragility has gained substantial interest from both the public and academia, particularly in complex systems and risk engineering communities \citep{aven_new_2014, aven_implications_2015, thekdi_integrated_2019, aven_risk_2022, axenie_antifragility_2024}. However, the design principles for realizing antifragile transportation systems and control are generally unexplored. One promising approach to induce antifragility is to leverage learning-based algorithms, such as Reinforcement Learning (RL), since RL agents can gradually adjust their decision-making when deployed to an environment subject to variations.

\begin{figure}[hbt!]
  \includegraphics[width=0.9\textwidth]{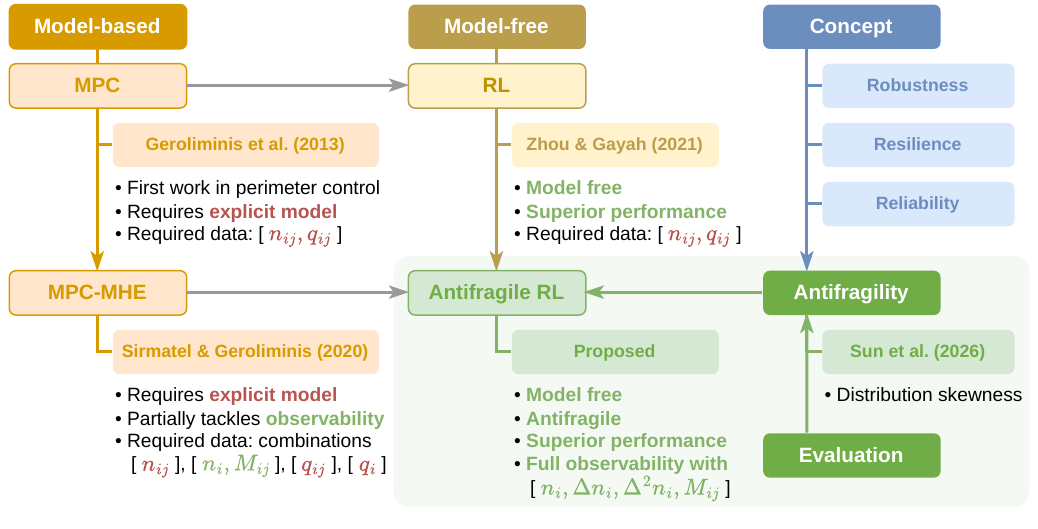}
  \centering
  \caption{Map of contributions.}
  \label{fig: contribution}
\end{figure}

The primary goal of this paper is to develop an antifragile perimeter control algorithm capable of adapting to escalating disruptions stemming from urban densification and the increasing frequency of natural hazards over time. Fig. \ref{fig: contribution} presents a map of contributions while highlighting the key references. The key contributions of this research can be summarized as follows: 

\begin{enumerate}
    \item \textit{Conceptual contribution:} We formalize the concept of antifragility as opposed to other related and commonly used terms in the risk engineering and transportation domains.
    \item \textit{Methodological contribution:} We design how antifragility can be incorporated into RL algorithms to achieve superior performance compared to benchmark methods, tackling both fragile performance and observability issues.
    \item \textit{Evaluatory contribution:} We introduce a skewness-based quantitative indicator to showcase the antifragile properties of our proposed algorithm under increasing disruptions.
    \item \textit{Model-agnostic applicability:} We demonstrate the effectiveness of the proposed antifragile module on other state-of-the-art RL algorithms and with a real-world case study.
\end{enumerate}

The remainder of this paper is structured as follows. In Section \ref{sec: literature}, we introduce relevant literature on multiple aspects covered in this work, while Section \ref{sec: problem formulation} mathematically formulates the cordon-shaped perimeter control simulation environment. Methodologies related to incorporating antifragility into RL algorithms are detailed in Section \ref{sec: methodology}. Section \ref{sec: experiment} discusses the simulation setups and parametrization. Results are presented in Section \ref{sec: results}, followed by concluding remarks and further discussions in Section \ref{sec: conclusion}. 

\section{Literature review}
\label{sec: literature}

This section reviews the relevant literature across three topics intertwined in this work. First, a macroscopic traffic model and the control strategy applied in this work are introduced, which serve as the basis of model dynamics for the simulation environment. Next, we review how robustness, resilience, or reliability can be induced with RL algorithms to counter real-world model uncertainty. Lastly, these terminologies related to antifragility are distinguished while we introduce the design philosophy of an antifragile system.

\subsection{Macroscopic fundamental diagram and perimeter control}

Alleviating urban network congestion can be realized through various traffic control strategies. Ever since \cite{geroliminis_existence_2008} demonstrated the existence of the Macroscopic Fundamental Diagram (MFD) with empirical data, the mathematical relationship between flow-density-speed has paved the path for the development of control policies at a macroscopic level. This breakthrough enabled computationally feasible real-time traffic management for large-scale networks, such as perimeter control \citep{keyvan-ekbatani_exploiting_2012, geroliminis_optimal_2013, kouvelas_enhancing_2017}, congestion pricing \citep{genser_dynamic_2022}, and route guidance \citep{fu_perimeter_2022}. 

Perimeter control is among the strategies that have attracted immense attention and research effort. By restricting incoming flows from adjacent regions into a protected zone, perimeter control maintains the traffic density in the protected area below the critical density for satisfactory serviceability. \cite{geroliminis_optimal_2013} proposed an optimal perimeter control method using Model Predictive Control (MPC) and proved its effectiveness compared to a greedy controller in a cordon network. Later, \cite{sirmatel_nonlinear_2020} introduced a Moving Horizon Estimation (MHE) scheme together with MPC to tackle measurement noise and observability issues. One major issue of the previous works is the assumption of MFD homogeneity. Consequently, substantial effort has been devoted to investigating partitioning algorithms so that a well-defined MFD can be created for individual sub-networks, as in \cite{ambuhl_approximative_2019}, creating a multi-reservoir system for applying perimeter control \citep{aboudolas_perimeter_2013}. However, MFDs in the real world can hardly be well-defined, as demonstrated in \cite{ambuhl_disentangling_2021} with loop detector data over a year. \cite{WANG2015} also demonstrated that adverse weather conditions and traffic incidents can alter the shape of the MFDs, while recovery from the peak-hour congestion frequently leads to hysteresis \citep{gayah_clockwise_2011}. These phenomena could potentially violate the mathematical model that serves as the foundation for the established model-based perimeter controllers. Therefore, \cite{zhou_model-free_2021} introduced an RL-based algorithm that exhibits superior performance and offers the advantage of being model-free.

\subsection{Leveraging RL to induce robustness, resilience, and reliability}
\label{sec: leveraging}

Similar to traffic networks and system control, real-world systems of all kinds can be subject to disruptions. As RL offers flexibility in defining state representations, actions, and rewards, researchers have leveraged such features to enhance system performance under adverse events. Here, we present a review of studies that apply RL in reliability engineering and traffic engineering domains, claiming enhancements in robustness, resilience, reliability, or related properties. 

Most existing studies on data-driven maintenance scheduling algorithms for infrastructure systems or mechanical equipment prioritize costs or efficiency, often lacking an orientation to reliability or resilience. To bridge such gaps, \cite{yuan_framework_2025} proposed an innovative reliability reward component based on the Gamma-Wiener process, enabling the RL agent to achieve a balance between dynamic reliability and maintenance cost of mechanical equipment. Likewise, \cite{liang_resilience-based_2026} proposed a specialized resilience metric, which encourages the RL agent to accelerate post-disaster recovery while explicitly considering system resilience. Furthermore, \cite{yin_variational_2025} coupled the DistFlow model for feasible resilience assessment and Bayesian learning to quantify system uncertainties and promote resilient post-disruption recovery. Risk and uncertainty can also be directly modeled in the reward function as in \cite{chiou_learning_2024}. On top of reward shaping, \cite{xu_enhancing_2026} also introduced a phase-adaptive discount factor to enhance the resilience of combat system-of-systems. 

In traffic networks, RL algorithms can also be engineered to strengthen the robustness or resilience of different control systems, such as traffic light control and perimeter control. For instance, \cite{rodrigues_towards_2019} induced robustness by adding the elapsed time since the last green signal for each phase, \cite{tan_robust_2020} experimented with speed or residual queue as an additional state representation in the RL algorithm, \cite{chu_multi-agent_2020} supplemented the control policies of neighboring intersections as additional information to the agents, and \cite{zhou_scalable_2023} used an extra binary congestion indicator in the state space. As a result, the algorithms are provided with additional information related to disturbances or disruptions. 

The analysis of the state-of-the-art RL studies considers reversals or sudden changes in the state-action-reward dynamics, which evoke unanticipated uncertainty. The problem in these contexts is often to respond to unexpected results appropriately, since they might indicate a shift in the environment. In this case, exploration refers to the process of searching for new information to improve the RL agent's understanding of the traffic dynamics under disruptions, which would then be used to identify better courses of action. By exploring how robustness, resilience, and reliability can be achieved using RL-based traffic control algorithms, we can potentially develop antifragile traffic control systems through similar approaches.

\subsection{Antifragility: definitions and distinctions}
\label{sec: antifragility}

Before detailing antifragile system design, we first need to review the distinctions between antifragility and other closely related terms, which include robustness, resilience, reliability, stability, and adaptiveness. Following the definitions proposed in \cite{zhou_resilience_2019}, robustness is concerned with assessing a system's capacity to preserve its initial state and resist performance deterioration under minor disturbances, while resilience emphasizes the ability and speed of a system to recover from major disruptions to its original state. Reliability in transportation has a much broader range of meanings. \cite{pennetti_evaluating_2020} proposed travel time reliability, which is conceptually similar to robustness in measuring deviation from normal operation, but focuses on the probability of such deviation exceeding a certain threshold. \cite{cats_robustness_2017} introduced a reliability assessment indicator to account for performance loss for the entire range of possible capacity reductions. Stability can also bear multiple definitions based on the context, such as describing the closed-loop stability of system control as in \cite{sirmatel_stabilization_2021} or MFD hysteresis based on the traffic flow theory \citep{gayah_clockwise_2011}. Adaptiveness is the capacity of a system to modify its own characteristics to maintain autonomous function across diverse environments \citep{hooker2011introduction}, which aligns with the concept of proto-antifragility proposed in \cite{taleb2012antifragile}. However, adaptiveness does not necessarily guarantee a consistent performance improvement under growing magnitudes of disruptions. Therefore, \cite{taleb2012antifragile} introduced the concept of antifragility, emphasizing the concave response of the system under increasing disruptions, which can be mathematically formulated with Jensen's inequality $\mathbb{E}[g(X)] \leq g(\mathbb{E}[X])$. It should be noted that while (anti-)fragility can be an innate property of a system, proper intervention and control strategies can lead the system to be more antifragile \citep{axenie_antifragility_2024}. For instance, building on the mathematical proof in \cite{sun_fragile_2024} that urban road networks are naturally fragile, this paper seeks to develop antifragile solutions that mitigate such fragility through perimeter control.

Ever since the concept of antifragility was proposed, it has garnered increasing attention across various disciplines in recent years, such as energy \citep{coppitters_optimizing_2023}, electricity \citep{rachunok_sensitivity_2020}, biology \citep{kim_antifragility_2020}, medicine \citep{axenie_antifragile_2022}, cyber-systems \citep{chatterjee_iterative_2020}, robotics \citep{axenie_antifragile_2023}, and machine learning \citep{pravin_fragility_2024}. Researchers have also proposed methods to promote the antifragile property of a system by emphasizing the derivatives of system states to capture the temporal evolution patterns of the system dynamics, i.e., how fast the system state deviates toward possible black swan events and the curvature of this deviation \citep{taleb_mathematical_2013, axenie_antifragile_2022}. With this additional information, the system can potentially anticipate ongoing disruptions and become more proactive in facing drastic changes. Similar to its function in resilience \citep{kamalahmadi_impact_2022}, redundancy can also be added to the system to induce antifragility \citep{munoz_resilience_2022}. Other feasible approaches include time-scale separation and attractor dynamics \citep{axenie_antifragile_2022}. However, leveraging the potential of antifragile system design for the operation of transportation networks remains an unexplored notion.

\section{Problem formulation}
\label{sec: problem formulation}

\begin{figure}[hbt]
  \centering
    \subfigure[The cordon-shaped urban network]{%
      {\includegraphics[width=0.45\textwidth, trim=0 5 0 10]{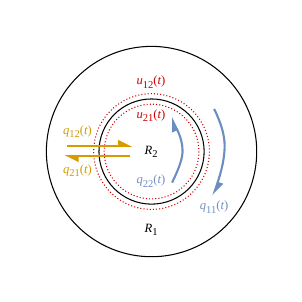}}\label{fig: corden}}
    \subfigure[MFDs for the inner and outer regions.]{%
      {\includegraphics[width=0.48\textwidth]{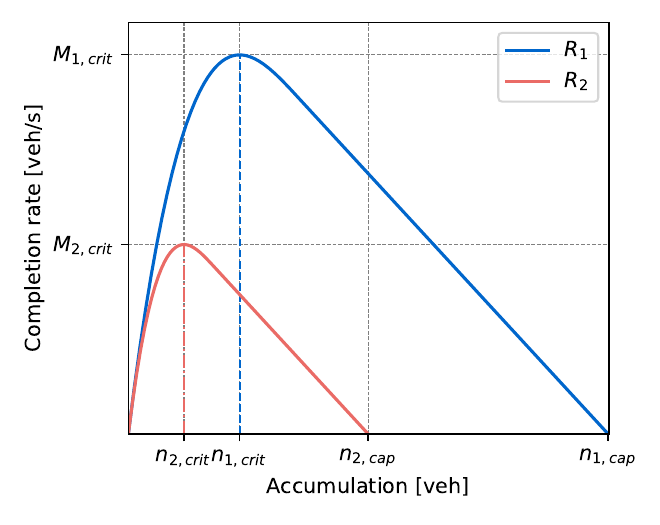}}\label{fig: MFD}}
    \caption{The network structure and the related MFDs.} 
\end{figure}

This paper studies perimeter control between two homogeneous cordon-shaped urban networks, as in \cite{geroliminis_optimal_2013}, with the inner region representing a city center, as shown in Fig. \ref{fig: corden}. The total number of vehicles in region $i$ at time $t$ is denoted as $n_i(t)$, while the Origin-Destination (OD) from region $i$ to region $j$ can be further divided as $n_{ij}(t)$. The inner and outer regions are assumed to have different MFDs to represent the capacity difference of accommodating vehicles within the network between the city center and the surrounding region, which are defined as $G_i(n_i(t))$ as illustrated in Fig. \ref{fig: MFD}. The total trip completion rate $M_i(t)$ for region $i$ at time $t$ can be determined through the corresponding MFD, which comprises both the intraregional trip completion $M_{ii}(t)$ and the interregional transfer flow $M_{ij}(t) \: (i\neq j)$ with $i, j \in \{1,2\}$. While $q_{i}(t)$ denotes the new demand within region $i$ at time $t$, $q_{ij}(t)$ represents the additional demand based on the OD. To protect a certain region from being overflown by high traffic demand, the percentage of the transfer flow $M_{ij}(t)$ allowed to pass across the boundary is regulated by the perimeter controllers, denoted as $u_{ij}(t) \: (i\neq j)$ with $i, j \in \{1,2\}$. All notations applied in this work, including those in the subsequent sections, are summarized in Table \ref{tab: notation}.

\begin{table}[hbt!]
\small
\centering
\caption{List of notations}
\label{tab: notation}
\renewcommand{\arraystretch}{1.15}
\begin{tabular}{ll}
\hline
Symbol   & Description                                                            
\\ \hline
\multicolumn{2}{l}{1. Notations for the problem formulation (continuous form)}                      
\\ \hline
$t$      & Current time                                                          
\\
$\Delta t$      & Time step interval                                                            \\
$T$  &  Total simulation time                                               
\\ 
$n_{ij}(t)$ & Vehicle accumulation with OD from region $i$ to $j$ at time $t$              
\\
$n_i(t)$    & Vehicle accumulation in region $i$ at time $t$      
\\
$u_{ij}(t)$ & Perimeter control variables regulating flow from region $i$ to $j$ at time $t$  
\\
$q_{ij}(t)$ & Traffic demand with OD pair $i$ and $j$ at time $t$                           
\\
$G_i(n_i(t))$   &   Sum of trip completion and transfer flow in region $i$  at time $t$  
\\
$M_{ij}(t)$ & Trip completion with OD from region $i$ to $j$ at time $t$    
\\
$n_{i,\mathrm{cap}}$ & Maximal number of vehicles (jam accumulation) in region $i$
\\
$n_{i,\mathrm{crit}}$ & Vehicle accumulation with highest completion rate in region $i$      
\\ 
$J$      &  Objective function                                                  
\\ \hline
\multicolumn{2}{l}{2. Notations in RL (discrete form)}             
\\ \hline
$k$      & Current time step, $k=\lfloor t/\Delta t\rfloor$                         
\\
$K$      & Total time steps, $K=T/\Delta t$ 
\\
$\mathcal{S}$ & State space, the whole set of states the RL agent can transition to 
\\ 
$s_k$ & $s_k \in \mathcal{S}$, the observable state in simulation at timestep $k$ 
\\ 
$\mathcal{A}$ &  Action space, the whole set of actions the RL agent can act out
\\ 
$a_k$ & $a_k \in \mathcal{A}$, the action taken in simulation at timestep $k$  
\\ 
$\mathcal{R}$ &  The reward function for the RL agent 
\\ 
$r_k$ & $r_k = \mathcal{R}(s_k, a_k)$, the received reward with state $s_k$ and action $a_k$ at timestep $k$
\\  
$\gamma$ & Discount factor to favor rewards in the near future 
\\ 
$Q(s_k,a_k)$ & Expected long-term return for taking action $a_k$ in state $s_k$ at timestep $k$ 
\\ 
$\theta^{\mu}$ & Weight parameter of the deep neural network for the actor network 
\\ 
$\theta^{Q}$ & Weight parameter of the deep neural network for the critic network 
\\ 
$y_i$ & Expected long-term return calculated with the target critic network  
\\ 
$L$ & The loss of the critic network 
\\ 
$\rho^{\beta}$ & All possible trajectories of $s_k$ 
\\
$I$ & The objective function for the actor-network 
\\ \hline
\multicolumn{2}{l}{3. Notations for the antifragile terms in RL (discrete form)}             
\\ \hline
$r_{\mathrm{com}, k}$ & Reward term in RL based on trip completion 
\\ 
$r_{\mathrm{red}, k}$ & Additional reward term in RL based on derivatives and redundancy 
\\ 
$r_{\mathrm{dam}, k}$ & Additional damping term to counter possible action oscillation 
\\ 
$\omega_{h}$ & The weight of the first derivative in the additional reward term $r_{\mathrm{red}, k}$  
\\ 
$\omega_{\Delta h}$ & The weight of the second derivative in the additional reward term $r_{\mathrm{red}, k}$  
\\ 
$\alpha_{i, k}$ & Binary variable determining the term to be a reward/penalty
\\ 
$h_{i, k}$ & The first derivative of traffic state at timestep $k$ 
\\ 
$\Delta h_{i, k}$ & The second derivative of traffic state at timestep $k$ 
\\\hline
\end{tabular}
\end{table}

Eq. \ref{eq: n_ii} describes the change rate of the intraregional vehicle accumulation of region $i$. It is the sum of the traffic demand in this region $q_{ii}(t)$ and the perimeter control regulated transfer flow  $u_{ji}(t)\cdot M_{ji}(t)$, then deducted by trip completion $M_{ii}(t)$. Likewise, the change rate of interregional traffic accumulation, as Eq. \ref{eq: n_ij} shows, is the difference between the interregional traffic demand $q_{ij}(t)$ and the regulated transfer flow $u_{ij}(t)\cdot M_{ij}(t)$:

\begin{subequations}
\begin{align}
\label{eq: n_ii}
\frac{dn_{ii}(t)}{dt} &= q_{ii}(t)+u_{ji}(t)\cdot M_{ji}(t)-M_{ii}(t) 
\\
\label{eq: n_ij}
\frac{dn_{ij}(t)}{dt} &= q_{ij}(t)-u_{ij}(t)\cdot M_{ij}(t)
\end{align}
\end{subequations}

The total trip completion $M_i(t)$ for region $i$ at time $t$ is calculated based on the trip accumulation and the related MFD $G_i(n_i(t))$, which is the sum of the intraregional trip completion $M_{ii}(t)$ and the interregional transfer flow $M_{ij}(t)$:

\begin{subequations}
\begin{align}
\label{eq: M_ii}
M_{ii}(t) &= \frac{n_{ii}(t)}{n_i(t)}\cdot G_i(n_i(t))
\\
\label{eq: M_ij}
M_{ij}(t) &= \frac{n_{ij}(t)}{n_i(t)}\cdot G_i(n_i(t))
\\
n_{i}(t) &= \sum_{j=1,2}n_{ij}(t)
\end{align}
\end{subequations}

The objective function is to maximize the throughput of this cordon-shaped network. This is defined as the sum of the intraregional trip completion $M_{ii}(t)$ in both the inner and the outer regions, while the interregional trip completion represents only vehicles finishing part of their trips.

\begin{subequations}
\label{eq: J}
\begin{equation}
J = \max\limits_{u_{ij}(t)}{\int_{0}^{T}\sum_{i=1,2}M_{ii}(t)dt}
\tag{\ref{eq: J}}
\end{equation}
\begin{align}
\textrm{s.t.}\quad & n_{ij}(t) \geq 0
\\
& n_{i}(t) \leq n_{i,\mathrm{cap}} 
\\
\label{eq: constraints}
& u_{\min} \leq u_{ij}(t) \leq u_{\max} 
\end{align}
\end{subequations}

Intraregional and interregional vehicle accumulation $n_{ii}(t)$ and $n_{ij}(t)$ are non-negative values, and $n_{i,\mathrm{cap}}$ is the maximal possible number of vehicles accumulated in region $i$, at which value a gridlock will occur in the network. $u_{\min}$ and $u_{\max}$ in Eq. \ref{eq: constraints} represent the lower and upper limits for the perimeter control variable $u_{ij}(t)$ for both directions. Such boundaries are in line with \cite{geroliminis_optimal_2013, zhou_model-free_2021}, as perimeter control is normally implemented through signalization. While $u_{\max}$ accounts for the lost time between the red and green phases, $u_{\min}$ is essential since an indefinitely long red light is rare in real-world urban networks.

In contrast to the objective function to be applied for control-based strategies, the reward function for the proposed antifragile RL-based algorithms $J_{af}$ is described in Eq. \ref{eq: J_RL} in discrete form, with $r_{\mathrm{com}, k}$ standing for the total trip completion in both regions at timestep $k$, that is $r_{\mathrm{com}, k}= \sum\limits^{k\Delta t+\Delta t}_{t=k\Delta t}\sum\limits_{i=1,2}M_{ii}(t)$.

\begin{equation}
\label{eq: J_RL}
J_{af}=\max\limits_{u_{ij}(k)}{\sum_{k=1}^{K}\left(r_{\mathrm{com}, k}+r_{\mathrm{dam}, k}+r_{\mathrm{red}, k}\right)}
\end{equation}

The second term $r_{\mathrm{dam}, k}$ is the damping term to counter the oscillating behavior of the RL agents due to the incorporation of derivatives in RL. The third term $r_{\mathrm{red}, k}$ represents the redundancy term that establishes a proper redundancy so that the proposed RL algorithm does not reward the agent for targeting the optimal critical accumulation. More explanation of the damping term $r_{\mathrm{dam}, k}$ and the redundancy term $r_{\mathrm{red}, k}$ can be found in Section \ref{sec: antifragile reward term}. 

\section{Methodology}
\label{sec: methodology}

In this section, we first discuss the methodologies of the benchmarks and use their benefits and drawbacks to motivate the development of our proposed antifragile perimeter control algorithm.

\subsection{Model predictive control and moving horizon estimation}
\label{sec: mpc}

Model Predictive Control (MPC) is a widely recognized control approach extensively used for regulating dynamic systems across various engineering fields, particularly in transportation and perimeter control. Readers interested in MPC and its applications in traffic engineering can refer to \cite{geroliminis_optimal_2013}. The MPC toolkit applied in this paper is introduced in \cite{lucia_rapid_2017}.

In \cite{sirmatel_nonlinear_2020}, Moving Horizon Estimation (MHE) is introduced to form an MPC-MHE framework for perimeter control. MHE enables real-time estimation of internal system states by solving an optimization problem that minimizes the discrepancy between predicted outputs and noisy sensor measurements. \cite{sirmatel_nonlinear_2020} further addresses the data observability issues in \cite{geroliminis_optimal_2013} as illustrated in Fig. \ref{fig: contribution}, i.e., the information on both the OD accumulation $n_{ij}(t)$ and the newly generated demand $q_{ij}(t)$ is almost impossible to acquire in real-time. Therefore, four different combinations of data availability were tested: one from either $n_{ij}(t)$ or $[n_{i}(t), M_{ij}(t)]$ and another one from either $q_{ij}(t)$ or $q_{i}(t)$. The accumulation $n_{i}(t)$ for region $i$ can be approximated with loop detector measurements, while the interregional trip completion $M_{ij}(t)$ can be directly counted through detectors at the intersections forming the perimeter border. The authors also claim $q_{i}(t)$ may be available under certain circumstances, such as when a substantial percentage of GPS information is readily accessible. However, due to the nature of the prediction horizon in MPC-MHE, such data has to be made available in both real time and the near future, rendering it impossible to collect, especially under disruptive conditions. We will further tackle such observability issues with the proposed antifragile perimeter control algorithm in the following Section \ref{sec: antifragile reward term}.

\subsection{The DDPG algorithm}
\label{sec: ddpg}

\begin{figure}[hbt!]
\includegraphics[width=0.75\textwidth]{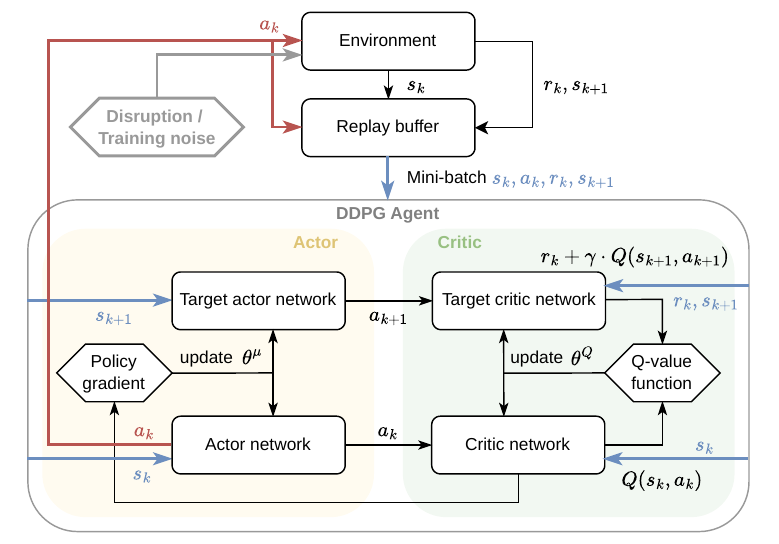}
\centering
\caption{DDPG scheme.}
\label{fig: DDPG}
\end{figure}

In RL algorithms, an agent or multiple agents interact with a preset environment and improve its decision-making capabilities, defined as action $a_k$ in an action space $\mathcal{A}$, based on the observable state $s_k$ in the state space $\mathcal{S}$ and the reward, defined as $r_k = \mathcal{R}(s_k, a_k)$, where $\mathcal{R}$ is the reward function. The improvement of decision-making is commonly realized through a deep neural network as a function approximator. The RL algorithm applied in this work is Deep Deterministic Policy Gradient (DDPG), as proposed in \cite{lillicrap_continuous_2019}. By applying an actor-critic scheme, DDPG can manage a continuous action space instead of a discrete action space. The DDPG algorithm can be divided into two main components, namely the actor and the critic, which are updated at each step through policy gradient and Q-value, respectively. The scheme of the DDPG algorithm applied in this paper is schematically illustrated in Fig. \ref{fig: DDPG}.

The actor-network is represented by $\mu(\cdot)$, and it determines the best action $a_k$ for the perimeter controller based on the current state $s_k$. The critic network $Q(\cdot)$ is responsible for evaluating whether a specific state-action pair at time step $k$ yields the maximal possible discounted future reward $Q(s_k, a_k)$. A common technique used in DDPG is to create a target actor network $\mu '(\cdot)$ and a target critic network $Q '(\cdot)$, which are a copy of the original actor and critic network but updated posteriorly to stabilize the training process and prevent overfitting, with the target maximal discounted future reward calculated as: 

\begin{equation}
\label{eq: y_i}
y_k = r_k + \gamma Q' (s_{k+1}, \mu'(s_{k+1}|\theta^{\mu'} )|\theta^{Q'})
\end{equation}

The critic network can then be updated by computing the temporal difference between the predicted reward as well as the target reward and minimizing the loss for a mini-batch $N$ sampled from the replay buffer:

\begin{equation}
L = \frac{1}{N} \sum_k (y_k - Q(s_k,a_k|\theta^Q))^2
\end{equation}

Afterward, the actor network can be updated with the sampled deterministic policy gradient: 

\begin{subequations}
\begin{align}
I =& \mathbb{E}_{s_k\sim \rho^{\beta}} 
[r(s,\mu (s|\theta^{\mu}))|_{s=s_k}]
\\[5pt]
\nabla_{\theta^{\mu}}I =& \mathbb{E}_{s_k\sim \rho^{\beta}} 
[\nabla_a Q(s,a|\theta^Q)|_{s=s_k,a=\mu(s_k)} \nabla_{\theta_{\mu}} \mu (s|\theta^{\mu})|_{s=s_k}]
\\
\nabla_{\theta^{\mu}}I \approx& \frac{1}{N} \sum _k \nabla_a Q(s,a|\theta^Q)|_{s=s_k, a=\mu(s_k)} \nabla_{\theta^{\mu}} \mu (s| \theta^{\mu})|_{s_k}
\end{align}
\end{subequations}

First developed in \cite{horgan_distributed_2018}, a similar Ape-X architecture as applied in \cite{zhou_model-free_2021} is also adopted in this work, which allows for multiple generations of simulations during training to be included and centrally learned for the best policy. A decaying Gaussian noise is also introduced to encourage the DDPG agent to explore the state space.

\subsection{Antifragility and the antifragile terms in RL}
\label{sec: antifragile reward term}

As reviewed in Section \ref{sec: leveraging}, proper information can be integrated into the RL algorithms to induce robustness or resilience. Following the same idea, we incorporate antifragile derivatives and redundancy terms based on the methodologies reviewed in Section \ref{sec: antifragility}. First, the benchmark work in \cite{zhou_model-free_2021} establishes the state as $s_k=[n_{ij, k}, q_{ij, k}]$, with $q_{ij, k}$ regarded as an estimate of average daily traffic demand. However, real-time traffic demand $q_{ij, k}$ can hardly be acquired, particularly when faced with disruptions. Therefore, we propose a state representation incorporating both the first- and second-order derivatives of vehicle accumulation, which can be computed as the first and second differences from the vehicle accumulation $[\Delta n_{ij, k}, \Delta^2n_{ij, k}]$ between two consecutive time steps in the discrete form. Similar to \cite{sirmatel_nonlinear_2020}, we also tackle the data observability issue in this work, by integrating trip completion $M_{ij, k}$ and replacing OD-pair accumulation $n_{ij, k}$ with regional accumulation $n_{i, k}$ as well as the differences. It should be noted that $M_{ij, k}$ in \cite{sirmatel_nonlinear_2020} refers specifically to the interregional transfer flow $M_{ij, k}, (i \ne j)$. However, when considering a rather homogenous average trip length, the regional trip completion rate $M_{i}(t)$ is linearly correlated to the network average flow and can be approximated as in \citep{geroliminis_existence_2008, zhou_model-free_2021}. Therefore, the intraregional completion rate can be further computed as $M_{ii, k}=M_{i, k}-M_{ij, k}, (i \ne j)$. For fair comparison, we study two sets of states representing either idealized full observability or real-world limited observability in Section \ref{sec: full_obs} and \ref{sec: limited_obs}, respectively.

Idealized full observability:

\begin{equation}
\label{eq: full}
s_k=[n_{ij, k}, \Delta n_{ij, k}, \Delta^2 n_{ij, k}, M_{ij, k}]
\end{equation}

Real-world limited observability:

\begin{equation}
\label{eq: limited}
s_k=[n_{i, k}, \Delta n_{i, k}, \Delta^2 n_{i, k}, M_{ij, k}]
\end{equation}

The action $a_k \in \mathcal{A}$ is defined the same as the control variables $u_{ij}$ in Section \ref{sec: problem formulation}. For the reward term $r_k$, while \cite{zhou_model-free_2021} uses merely the completion rate, in our proposed algorithm, the reward is defined with additional two terms, i.e., the damping term $r_{\mathrm{dam}, k}$ and the redundancy term $r_{\mathrm{red}, k}$, as Eq. \ref{eq: J_RL} shows. When the first and second differences $[\Delta n_{ij, k}, \Delta^2n_{ij, k}]$ or $[\Delta n_{i, k}, \Delta^2n_{i, k}]$ are incorporated into the state space $\mathcal{S}$ of the algorithm, significant oscillations in the perimeter control variables can often be observed, especially under scenarios with demand uncertainties. In perimeter control systems, which utilize coordinated signals at regional boundaries, control oscillations can lead to rapid fluctuations in green splits between successive cycles. While such instability may have a negligible impact on overall traffic metrics, real-world signal operations necessitate high levels of stability for practical deployment. Therefore, a damping term $r_{\mathrm{dam}, k}$ is introduced into the reward function to penalize potential oscillatory actions. While the absolute difference between two consecutive actions is always below $1$, the constant $\xi_1$ bounds the maximal penalty, whereas $\xi_2$ determines how the penalty attenuates when the change of control variables becomes small, indicating a large $\xi_2$ only penalizes the agent when the oscillation is substantial. 

\begin{equation}
\label{eq: damping}
r_{\mathrm{dam}, k} = - \xi_1 \sum_{i,j \in \{1,2\}} |u_{ij, k}-u_{ij, k-1}| ^ {\xi_2}, \quad i \neq j \text{ and } k\neq 1
\end{equation}

The last term $r_{\mathrm{red}, k}$ in the reward function functions as an additional term to build up redundancy in the system, resembling the buffer times in train scheduling \citep{corman_automated_2018}. In contrast to the fixed buffer times prevalent in routinized railway scheduling, we introduce a dynamic redundancy term that incorporates both first- and second-order derivatives and regional accumulation differences. This formulation specifically accounts for the constraints of real-world limited observability. Here, we summarize $r_{\mathrm{red}, k}$ as the sum of two terms $r_{\mathrm{red}, k} = H_k + \Delta H_k$ in discrete form, with $H_k$ being an overall term representing the first difference, and $\Delta H_k$ representing the second difference, which can be further expanded as:

\begin{subequations}
\label{eq: redundancy}
\begin{align}
H_k =& \sum_{i=1,2}H_{i, k} = \sum_{i=1,2} \omega_h \cdot h_{i, k} \cdot \alpha_{i, k} \cdot f(n_{i, k}, n_{i,\mathrm{crit}}, n_{i,\mathrm{cap}})
\\
\Delta H_k =& \sum_{i=1,2}\Delta H_{i, k} = \sum_{i=1,2} \omega_{\Delta h}\cdot \Delta h_{i, k} \cdot f(n_{i, k}, n_{i,\mathrm{crit}}, n_{i,\mathrm{cap}})
\end{align}
\end{subequations}

In Eq. \ref{eq: redundancy}, $\omega_h$ and $\omega_{\Delta h}$ are introduced as the weight constants for the derivatives to regulate their impact on the reward $\mathcal{R}$. $h_{i, k}$ and $\Delta h_{i, k}$ are the first and second differences of the traffic states on the MFD, $h_{i, k}$ is defined as the difference of trip completion over vehicle accumulation at the end of a time step versus at the beginning of the same time step. The second difference $\Delta h_{i, k}$ is calculated as the difference between $h_{i, k}$ of two consecutive time steps:

\begin{subequations}
\begin{align}
\label{eq: h_i}
h_{i, k} = \frac{M_{i, k}\ -\ M_{i, k-1}}{n_{i, k}\ -\ n_{i, k-1}}
\\[5pt]
\label{eq: dh_i}
\Delta h_{i,k} = h_{i,k}-h_{i,k-1} 
\end{align}
\end{subequations}

The binary variable $\alpha_{i, k}$ in $H_k$ was designed to reward the agent when it moves towards the desired direction on the MFD. For instance, when the traffic state lies in the congested zone of the MFD, the gradient of any step will be negative. However, a penalty should be applied when vehicle accumulation gets larger, while rewarding the agent when accumulation decreases.

\begin{equation}
\alpha_{i, k}= 
\begin{cases}
    1,              & \text{if } n_{i,k}\geq n_{i,k-1},\\
    -1,             & \text{otherwise.}
\end{cases}
\end{equation}

Here, $f(n_{i,k}, n_{i,\mathrm{crit}}, n_{i,\mathrm{cap}})$ computes a reduction factor to constrain the impact of the $r_{\mathrm{red}, k}$ term when the accumulation is around the critical accumulation $n_{i,\mathrm{crit}}$, where the $r_{\mathrm{red}, k}$ term should have the greatest impact, which is realized through a modified trigonometric function. Other functions, such as normal distribution, should be valid for achieving the same goal as well.

\begin{equation}
f(n_{i,k}, n_{i,\mathrm{crit}}, n_{i,\mathrm{cap}}) = 
\begin{cases}
    {\Big(1+\cos\Big(-\pi \cdot \cfrac{n_{i,\mathrm{crit}}-n_{i,k}}{n_{i,\mathrm{crit}}}\Big)\Big)}/2,              & \text{if } n_{i,k}\geq n_{i,\mathrm{crit}},\\
    {\Big(1+\cos\Big(-\pi \cdot \cfrac{n_{i,k}-n_{i,\mathrm{crit}}}{n_{i,\mathrm{cap}}-n_{i,\mathrm{crit}}}\Big)\Big)/2},             & \text{otherwise.}
\end{cases}
\end{equation}

\begin{figure}[hbt!]
  \centering
    \subfigure[$H_k$ for the first difference $h_{i, k}$.]{%
      {\includegraphics[width=0.48\textwidth]{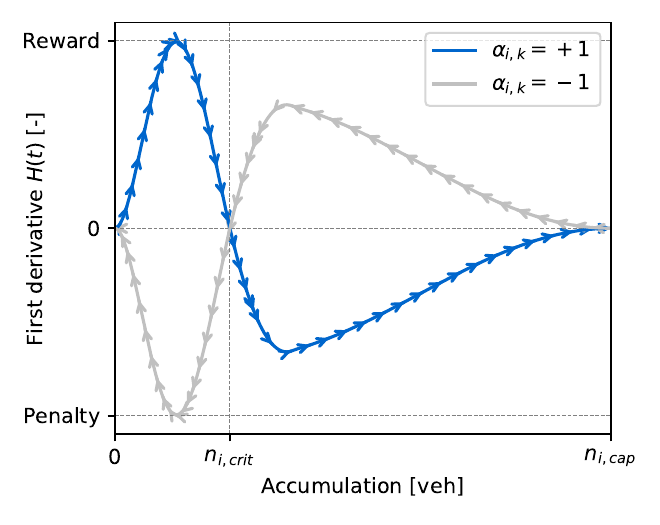}}
      \label{fig: h}}
    \subfigure[The impact on the MFD with $n_{i,k}\geq n_{i,k-1}$.]{%
      {\includegraphics[width=0.48\textwidth]{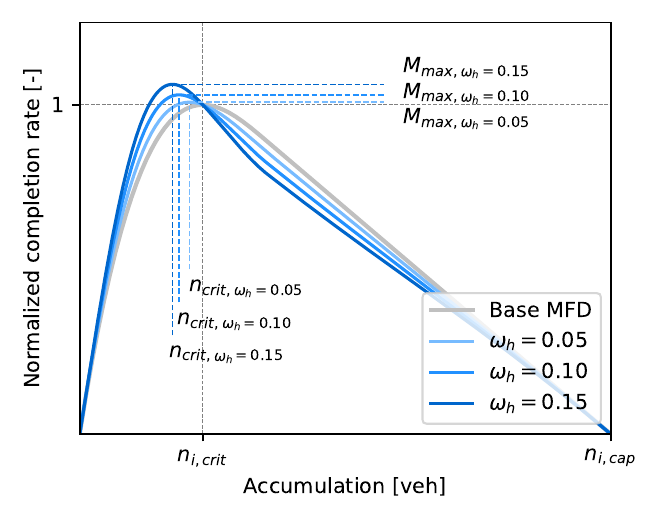}}
      \label{fig: MFD_mod}}
    \caption{Illustration of the term $H_k$ and its effect on the MFD.}
    \label{fig: h_mfd}
\end{figure}

\begin{figure}[hbt!]
  \includegraphics[width=0.85\textwidth]{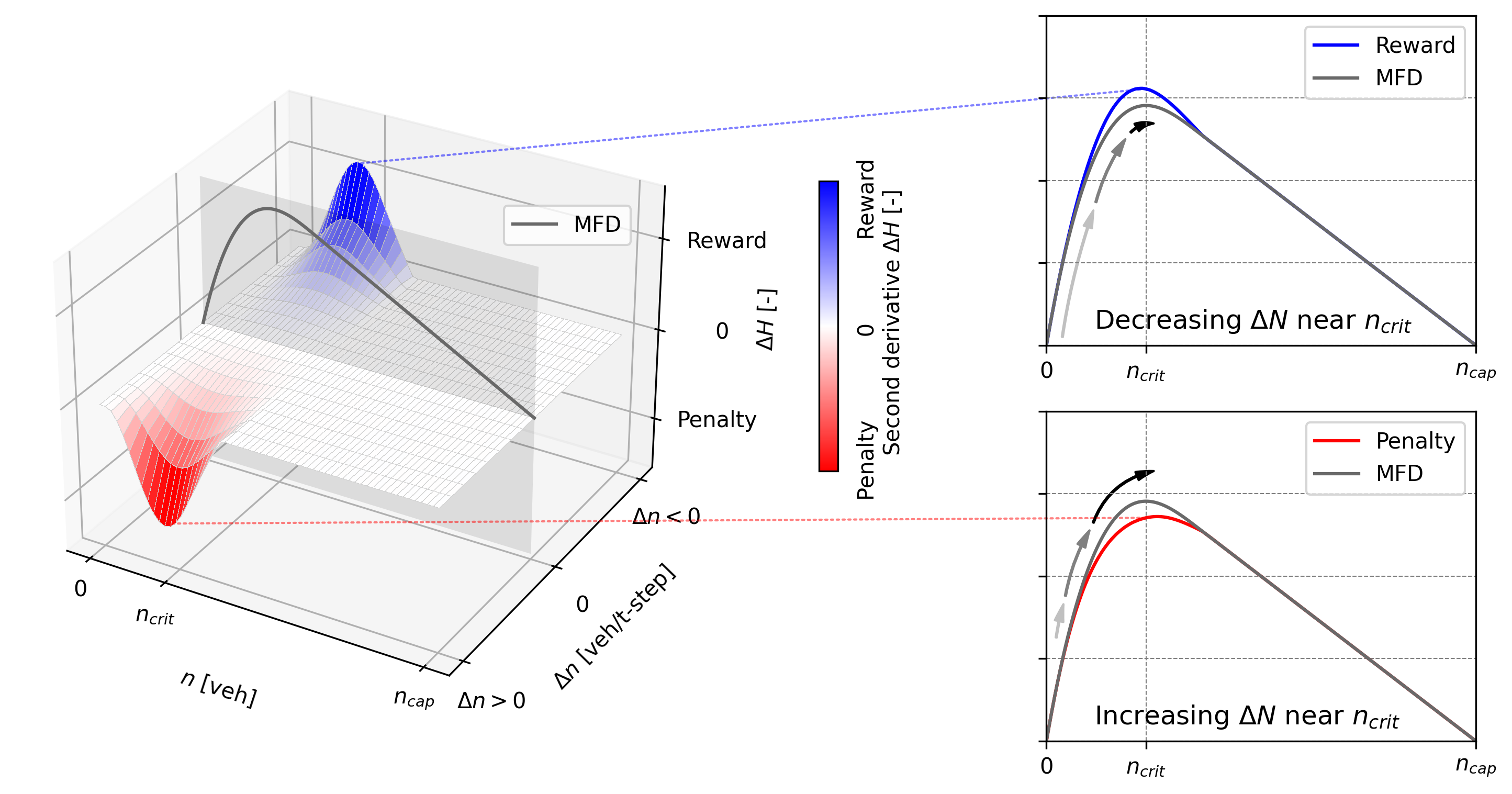}
  \centering
  \caption{Illustration of the term $\Delta H_k$.}
  \label{fig: dh}
\end{figure}

We illustrate the impact of $H_k$ and $\Delta H_k$ on building redundancy on the MFD in Fig. \ref{fig: h_mfd} and \ref{fig: dh}. The first difference $H_k$ in Fig. \ref{fig: h} rewards the agent when moving toward the critical accumulation $n_{i,\mathrm{crit}}$ to maximize trip completion. However, when $n_{i,k}$ approaches $n_{i,\mathrm{crit}}$, this term drops significantly and becomes a penalty when $n_{i,k}$ exceeds $n_{i,\mathrm{crit}}$. In Fig. \ref{fig: h_mfd}, we showcase the influence of this term on the MFD. With increasing weight coefficient $\omega_h$, $n_{i,\mathrm{crit}}$ of the modified MFD becomes lower compared to the original, and the reward decreases faster after the accumulation exceeds $n_{i,\mathrm{crit}}$. In this way, redundant overcompensation has been established to prevent accumulation from exceeding the critical accumulation when disruption happens unexpectedly.

The second difference $\Delta H_k$ is shown in Fig. \ref{fig: dh}. The y-axis represents how fast the traffic state changes, and the faster it reaches $n_{i,\mathrm{crit}}$, the greater the penalty will be. This observation is consistent with the redundant overcompensation and time-scale separation principles formalized in \cite{taleb_mathematical_2013} and practically applied in \cite{axenie_antifragile_2023}. On the contrary, a reward will be given if $n$ decelerates when approaching $n_{i,\mathrm{crit}}$. Likewise, $\Delta H_k$ is also dependent on the normalization factor $\omega_{\Delta h}$. With $H_k$ and $\Delta H_k$, the agent learns to be conservative when regulating the perimeter control variables to reach $n_{i,\mathrm{crit}}$. 

\section{Experiment application}
\label{sec: experiment}

Following the contribution map illustrated in Fig. \ref{fig: contribution}, we test our proposed antifragile perimeter control algorithm against three state-of-the-art perimeter control algorithms as benchmarks:

\begin{itemize}
  \item[-] MPC perimeter control proposed in \cite{geroliminis_optimal_2013}.
  \item[-] MPC-MHE perimeter control proposed in \cite{sirmatel_nonlinear_2020}.
  \item[-] Baseline RL perimeter control algorithm proposed in \cite{zhou_model-free_2021} with the original codebase\footnote[1]{\url{https://github.com/DongqinZhou/C-D-RL}}.
\end{itemize}

It is worth mentioning that this work employs MPC-MHE differently in comparison to \cite{sirmatel_nonlinear_2020}, where sensor measurement noise is assumed and addressed by MHE through state estimation. In contrast, we presume the presence of disruption but perfect sensor measurements as the real states. As a result, since there is no need to estimate the system states, MPC-MHE does not provide additional benefits over MPC alone in the context of demand disruptions. For supply disruptions, however, we introduce a disruption magnitude reduction coefficient $\eta$ as a state variable, and MPC-MHE can help estimate the coefficient and thereby the model shift, supposing such variation is well-defined and can be modeled.

\begin{figure}[hbt!]
  \includegraphics[width=1\textwidth]{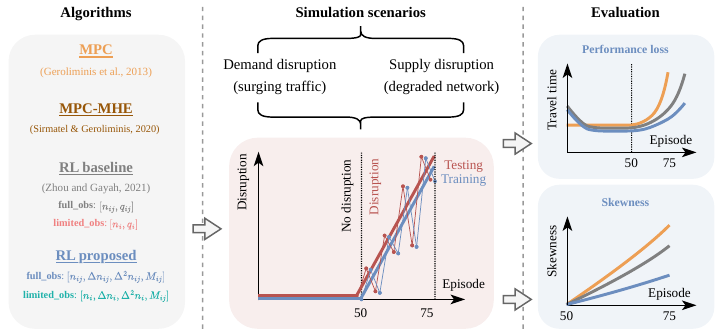}
  \centering
  \caption{Illustration of simulation scenarios and their evaluation.}
  \label{fig: structure}
\end{figure}

Since transportation systems inherently manage the imbalance between demand and supply, real-world disruptions can also be broadly categorized as either a demand or supply disruption. A demand disruption can be easily understood as a surging traffic demand due to social events or other occurrences. Population growth and urbanization can also be considered a long-term and gradually incremental demand disruption. On the other hand, a supply disruption can represent a network-level capacity drop due to adverse weather, which can be reflected by the decrease of the MFD profile, as in \cite{lu_traffic_2024}. In this paper, the performance of the investigated perimeter control algorithms is validated under both demand and supply disruptions, as illustrated in Fig. \ref{fig: structure}. A key feature of the experimental setup is the one-episode delay in training relative to the testing phase. In contrast to the standard training-testing paradigm in RL, we adopted an evaluation-first approach and employed a testing-training scheme. This design is intended to mirror the unpredictability of real-world disruptions. In addition to the deterministically linearly increasing magnitude of disruptions, as represented by the solid linear red line, we also tested the algorithms with uncertainties taken into account, as illustrated by the zigzag dotted red line, which will be explained in detail in the following section.  

\subsection{Simulation environment parametrization}
\label{sec: parametrization}

We simulate a cordon-shaped urban network with inner and outer regions represented by different MFDs, as Fig. \ref{fig: MFD} shows. These MFDs follow \cite{zhou_model-free_2021} and were originally approximated through the Yokohama loop detector dataset \citep{geroliminis_existence_2008}. However, since these MFDs are formulated as piecewise functions and are, although continuous, not differentiable, the lack of differentiability within MFDs can cause fluctuations when computing the first- and second-order differences and harm the efficacy of the redundancy term. Therefore, a minor modification has been made to slightly increase the gridlock accumulation by merely $3\%$, so that the whole MFD can be both continuous and differentiable. Other essential parameters, e.g., critical accumulation and maximal trip completion, remain the same as in \cite{zhou_model-free_2021}.

\begin{figure}[hbt!]
  \centering
    \subfigure[Base demand profile and demand disruption.]{%
      \resizebox*{8cm}{!}{\includegraphics{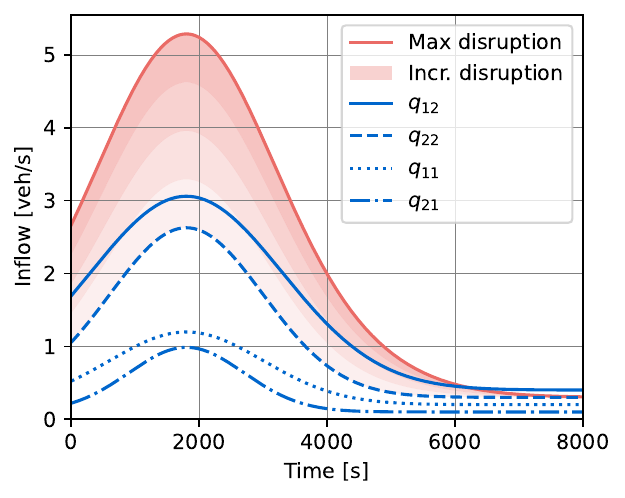}}
      \label{fig: inflow}}
    \subfigure[MFD and supply disruption for the inner region.]{%
      \resizebox*{7.8cm}{!}{\includegraphics{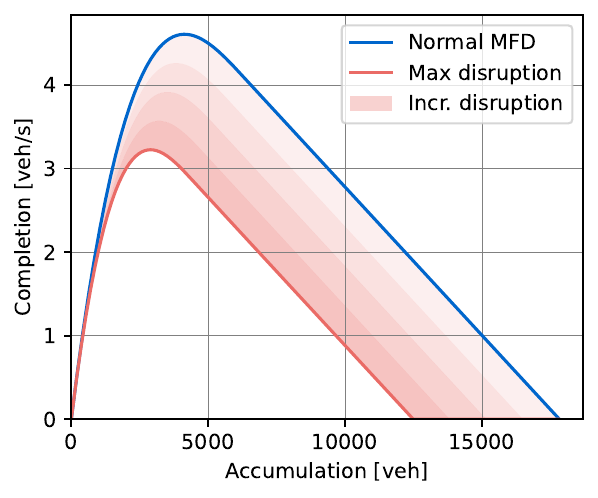}}
      \label{fig: supply}}
    \caption{Demand profile with or without surging demand.}
    \label{fig: Demand and supply disruptions}
\end{figure}

Traffic demand under no disruption is approximated based on \cite{geroliminis_existence_2008}. As the real-world peak hour traffic demand profile has more resemblance to a Gaussian distribution instead of being a simple trapezoidal shape \citep{mazloumi_using_2010}, the base demand $q_{ij}(t)=[q_{11}(t), q_{12}(t), q_{21}(t), q_{22}(t)]$ is illustrated as the blue curves in Fig. \ref{fig: inflow}. It consists of two components, which are a constant value $q_{ij,c}=[q_{11,c}, q_{12,c}, q_{21,c}, q_{22,c}]=[0.2, 0.4, 0.1, 0.3]$ in $\mathrm{veh/s}$ and a Gaussian term $q_{ij,G}(t)=C_{ij} \mathcal{N}_{ij}(\mu_{ij}, \sigma_{ij})$ with total number of vehicles following Gaussian as $C_{ij}=[3000, 10000, 2000, 7000]$ and the mean and standard deviation being $\mu_{ij}=[1800, 1800, 1800, 1800]$ and $\sigma_{ij}=[1200, 1500, 900, 1200]$ in seconds. Although real-world disruptions can occur in both the outer and inner regions, exerting a negative impact on overall performance, we focus on the critical scenarios where the potential of perimeter control can be better reflected. Since the inner region has a significantly smaller MFD profile compared to its outer counterpart, a surging disruptive traffic demand is presumed to be generated in the inner region. Numerically, $q_{22,G}$ is increased to simulate traffic from within the city center, reaching a total disruption demand of $12000$ vehicles at episode $75$, accounting for approximately one-third of the normal demand.

Supply disruptions can manifest in various forms, such as adverse weather conditions, traffic accidents, and road maintenance. However, limited research has been dedicated to understanding the exact correlation between such events and their impact on the MFD. In this work, we adopt a proportional decrease in both maximal accumulation and completion of the MFD in resemblance to \cite{lu_traffic_2024}, as shown in Fig. \ref{fig: supply}, representing lane closure events due to road construction works, traffic accidents, and natural disasters. Similar to the critical demand disruption on $q_{22,G}$, critical supply disruption requires the inner MFD being compromised with maximal $\eta=0.3$ while the outer MFD remains intact. Note that supply disruptions can be modeled in different ways for other events. For instance, considering the performance loss due to vehicle infrastructure interaction, \cite{ambuhl_functional_2020} proposed the concept of infrastructure potential $\lambda$ to represent how efficiently the network infrastructure is utilized, with a smaller $\lambda$ indicating the infrastructure is operated with higher efficiency.

The total simulation duration $T$ is 3 hours, and each time step $\Delta t$ takes $180$ seconds. The third and last hour has little demand and mainly serves as the unloading process. Another constraint is the lower and upper bounds for the perimeter control variable $u_{ij,k} \in [0.1, 0.9]$ as explained in Eq. \ref{eq: constraints}. The initial vehicle accumulation is set to be $n_{ij,0}=[600, 1300, 300, 2400]$ so that the accumulation remains approximately at an equilibrium at the beginning of the simulation. Each scenario is run $25$ times since randomness is inherent in RL and its performance may vary across simulation iterations. Each iteration lasts for $75$ episodes, with the first $50$ under no additional disruption so that the RL agent can be properly trained under the base demand profile with predefined training noise. Incremental magnitude of disruptions is introduced in the subsequent $25$ episodes, with the value deliberately set to be the same as the number of simulation iterations. Under disruption uncertainties, as illustrated by the zigzag lines for the simulation scenarios in Fig. \ref{fig: structure}, a list of multipliers $\epsilon=[\epsilon_1, \epsilon_2, \ldots, \epsilon_{24}, \epsilon_{25}]$ following a normal distribution $\epsilon \sim \mathcal{N} (1, 0.15)$ is randomly generated to introduce uncertainties by computing the element-wise multiplication with the list of disruption magnitudes. The same multiplier list is applied for all scenarios that are subject to disruption magnitude uncertainties. Also, it is cyclically shifted for each simulation iteration. For example, the list would be $\epsilon=[\epsilon_2, \epsilon_3, \ldots, \epsilon_{25}, \epsilon_1]$ for the second iteration. Through cyclical shifting, the disruption magnitudes from different episodes will experience exactly the same list of uncertainty multipliers, but with different sequences.

The most important hyperparameters for both the baseline and the proposed RL algorithms are in accordance with \citep{zhou_model-free_2021} and summarized in Table \ref{tab: hyperparameters}. Note that the minimal learning rate values are relatively larger than common RL algorithms, since we aim to get a trade-off between optimality and adaptiveness when the algorithm is experiencing disruptions. The number of simulations generated under the Ape-X architecture is $32$ per training episode. The coefficients $\xi_1$ and $\xi_2$ for the damping term $r_{\mathrm{dam}, k}$ in Eq. \ref{eq: damping} against oscillatory actions are set to be $1$ and $6$. The weights $\omega_h$ and $\omega_{\Delta h}$ for the redundancy term $r_{\mathrm{red}, k}$ are set to be $0.01$ and $0.02$. Both the prediction horizon and control horizon of MPC and MPC-MHE are set to $10$. 

\begin{table}[hbt!]
\footnotesize
\caption{List of hyperparameters.}
\label{tab: hyperparameters}
\centering
\begin{tabular}{L{4cm}C{2.5cm}C{2.5cm}}
\toprule
\textbf{Hyperparameter}     & \textbf{Value}             \\ \midrule
Replay buffer               & 10,000               &                       \\
Sample size                 & 1,000                &                       \\
Batch size                  & 256                  &                       \\
Discount factor             & 0.90                 &                       \\
\underline{{\textbf{DDPG}}}             &  &  \\
Noise initial scale  & 0.3                  &                       \\
Noise linear decay   & 0.003                &                       \\
Noise minimal scale  & 0.1                  &                       \\
Target network update       & 5                    &                       \\
\underline{{\textbf{TD3}}}             &  &  \\
Soft update  & 0.001                  &                       \\
Policy noise   & 0.1                &                       \\
Noise clip  & 0.2                  &                       \\
Policy delay       & 2                    &                       \\
\underline{{\textbf{SAC}}}             &  &  \\
Entropy coefficient  & \texttt{auto\_alpha}                  &                       \\
\texttt{log\_std}  clip  & [-20, 2]                  &                       \\
\underline{{\textbf{PPO}}}             &  &  \\
Entropy coefficient  & 0.01                  &                       \\
Clip ratio  & 0.2                  &                       \\
GAE Parameter  & 0.97                 &                       \\
PPO epoch  & 10                 &                       \\

\underline{\textbf{Learning rate}}             & \underline{\textbf{Actor}} & \underline{\textbf{Critic}} \\
Initial learning rate       & 0.004                & 0.008                 \\
Learning rate decay         & 0.98                 & 0.98                  \\
Minimal learning rate       & 0.0005               & 0.001                \\
Epoch                       & 2                    & 128                   \\ 
\underline{\textbf{Network architecture}}             & \underline{\textbf{Output}} & \underline{\textbf{Activation}} \\
First layer                 & 64                 & ReLU                 \\
Second layer                & 64                 & ReLU                  \\
Output layer                & 2                  & tanh                \\
\bottomrule
\end{tabular}
\end{table}

\subsection{Performance evaluation}
\label{sec: evaluation}

To showcase antifragility and the associated performance loss, the primary performance indicator evaluated in this work is the Total Time Spent (TTS), which is calculated by adding up the number of vehicles within the network at each second of the simulation. Although the reward in the RL algorithm is defined based on trip completion with additional damping and redundancy terms, two scenarios with the same trip completion at the end of the simulation may exhibit distinct TTS. And the one with the lower value should be regarded as having demonstrated a superior overall performance.

As \cite{sun_fragile_2024} has recently proved the fragile nature of urban road networks, fully antifragile traffic control strategies may be unattainable. When the magnitude of disruptions is sufficiently large, urban road networks will inevitably become fragile due to their real-world constraints. Therefore, this paper aims to demonstrate that the proposed perimeter control algorithm is relatively more antifragile than the state-of-the-art baseline algorithms. To quantify the antifragility of different algorithms, we calculate the distribution skewness based on the TTS sampled from the last $N_{\mathrm{episode}}=25$ incremental episodes, with $\mu_{\mathrm{TTS}}$ and $\sigma_{\mathrm{TTS}}$ denoting their mean and the standard deviation:

\begin{align}
    s=\frac{1}{N_{\mathrm{episode}}}\sum_{i=1}^{N_{\mathrm{episode}}}\left(\frac{\mathrm{TTS}-\mu_{\mathrm{TTS}}}{\sigma_{\mathrm{TTS}}}\right)^3
\end{align}

A skewness computed as $0$ indicates that the system itself or the applied algorithm makes the system neither fragile nor antifragile. Negative skewness indicates a longer or fatter left tail of the distribution and thus a higher degree of concavity in the performance function, which showcases antifragility. On the other hand, a positive skewness signifies fragility. For the detailed discussion regarding skewness and antifragility, interested readers may refer to \cite{sun_fragile_2024}. 

\section{Results}
\label{sec: results}

Following the simulation setup in Section \ref{sec: experiment}, we first evaluate the performance and antifragile characteristics of the studied approaches under idealized conditions with full observability. Next, we examine how the RL-based algorithms perform under real-world constraints with limited observability. Subsequently, the generalizability of the proposed method is validated by incorporating the antifragile module into other state-of-the-art RL algorithms. Lastly, we examine the performance of the investigated algorithms in a case study abstracted from a real-world network and disruptions. Each scenario is averaged and evaluated over 25 simulation runs.

\subsection{Performance under full observability}
\label{sec: full_obs}

Under idealized full observability, all algorithms are given perfect OD information $[n_{ij}(t),q_{ij}(t)]$ without uncertainties. Fig. \ref{fig: incr_d} presents the performance curves of the algorithms under the cordon-shaped perimeter control environment with demand disruptions, i.e., MPC in orange, baseline RL in gray, and the proposed antifragile RL algorithm in blue. The first 50 episodes represent the training process for the RL-based algorithms under the static demand profile in Fig. \ref{fig: inflow}. The TTS curves of the two RL-based algorithms gradually decrease and reach a comparable performance to MPC. The 50 episodes are then followed by another 25 episodes with linearly increasing demand disruption. The investigated algorithms demonstrate distinct capabilities of learning and adapting to disruptions and performance variance. The proposed antifragile RL algorithm exhibits both superior performance and reduced variance, indicated by its significantly narrower shaded area. Furthermore, the performance curve of the proposed method also appears less convex than those of the other algorithms. 

\afterpage{%
  \clearpage
    \begin{landscape} 
    \begin{figure}[htbp]
    \centering
    \subfigure[Performance curves in TTS.]{%
      {\includegraphics[width=0.485\textwidth]{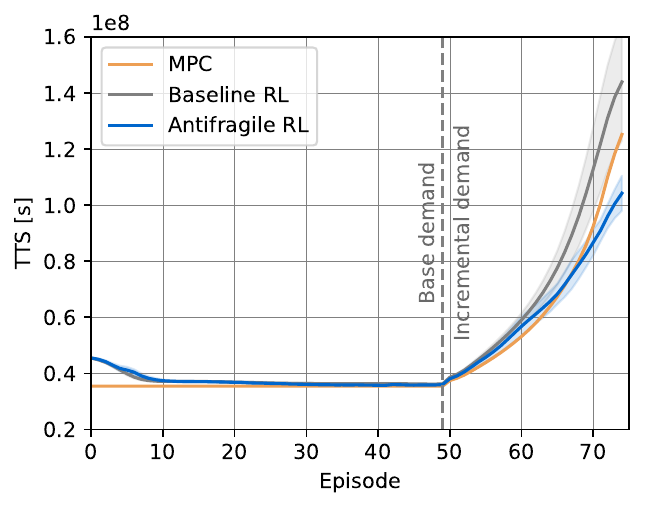}}
      \label{fig: incr_d}}
    \subfigure[Normalized over baseline RL.]{%
      {\includegraphics[width=0.43\textwidth]{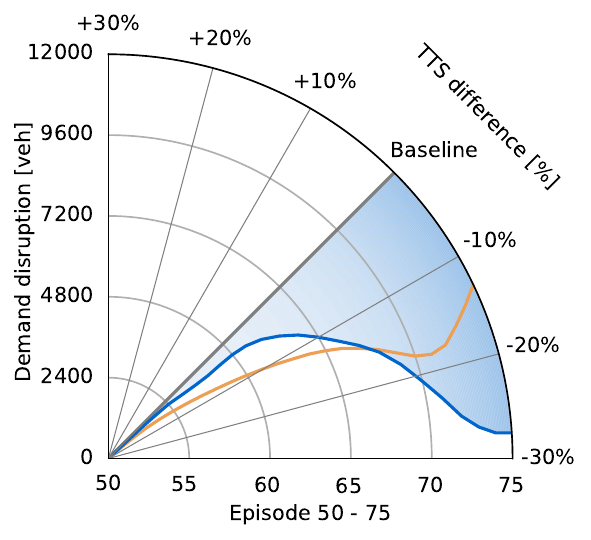}}
      \label{fig: incr_d_p}}
    \subfigure[Skewness up to the $n$-th episode.]{%
      {\includegraphics[width=0.369\textwidth]{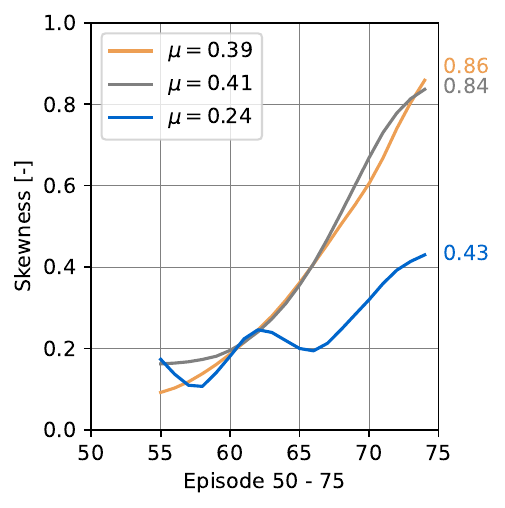}}
      \label{fig: incr_d_sk}}
    \caption{Performance curves under incremental demand disruptions.}
    \label{fig: d}

    \centering
    \subfigure[Performance curves in TTS.]{%
      {\includegraphics[width=0.485\textwidth]{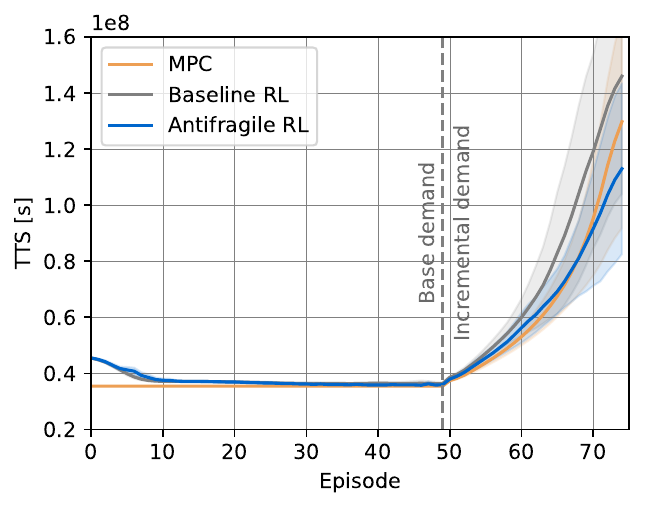}}
      \label{fig: incr_d_un}}
    \subfigure[Normalized over baseline RL.]{%
      {\includegraphics[width=0.43\textwidth]{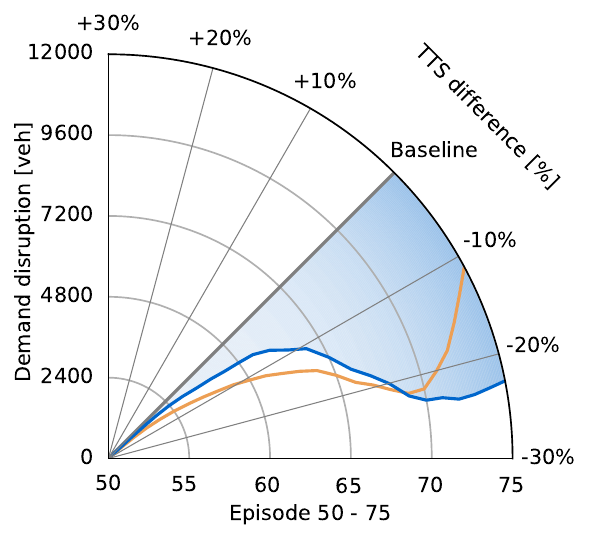}}
      \label{fig: incr_d_p_un}}
    \subfigure[Skewness up to the $n$-th episode.]{%
      {\includegraphics[width=0.369\textwidth]{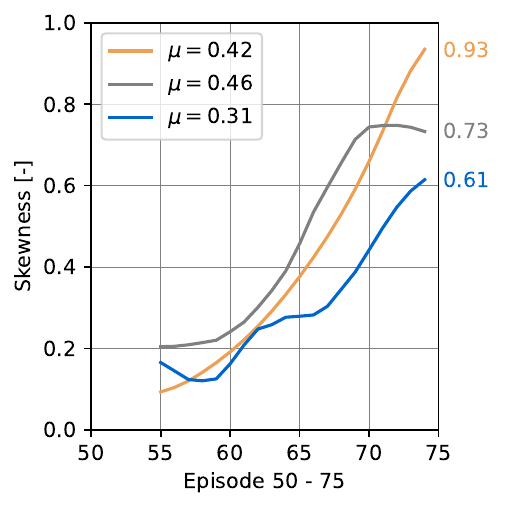}}
      \label{fig: incr_d_sk_un}}
    \caption{Performance curves under incremental demand disruptions with uncertainty.}
    \label{fig: d_un}
    \end{figure}
    \end{landscape}
  \clearpage
}

Such superior performance between episode $50-75$ under incremental demand is further quantitatively gauged with a polar plot in Fig. \ref{fig: incr_d_p}, which shows the performance difference in TTS reduction normalized over the baseline RL algorithm. Instead of variance, the shaded blue area highlights the performance gain of our proposed method compared to the baseline RL algorithm, showing a steady increase in TTS reduction and being largely concave. The average performance gain is $10.7\%$ and reaches $27.6\%$ at episode $75$ under incremental demand disruption. On the other hand, although MPC performs relatively better when the magnitude of disruption is low, its performance quickly converges toward the baseline RL under high demand disruptions, achieving a $13.0\%$ improvement at peak disruption. 

To quantify the antifragile properties of each perimeter control algorithm, we compute the distribution skewness for each method up to the $n$-th episode as well as the average skewness $\mu$ across all $25$ episodes under incremental disruptions, as in Fig. \ref{fig: incr_d_sk}. Note that due to the inherent variability of RL-based algorithms, as well as the high sensitivity of the distribution skewness on such variability, we apply a sliding window of 5 episodes to smooth the skewness curve. In addition, a buffer containing episodes $50-55$ is excluded to yield meaningful skewness values. The skewness of the baseline RL in gray rises rapidly to $0.84$, while MPC progresses to $0.86$. On the contrary, the skewness curve in blue of the antifragile RL-based algorithm achieves an ultimate skewness at $0.43$. 

Fig. \ref{fig: d_un} shows the performance when stochasticity is introduced into the magnitude of disruptions, following a disruption cyclic shifting scheme discussed in Section \ref{sec: parametrization}. The results display a consistent trend with the non-stochastic scenario, but with the performance curves less smooth, suggesting that the presence of uncertainty adds complexity to the training process. Although the performance of both RL-based methods deteriorates slightly, the proposed antifragile RL algorithm still outperforms the baseline RL by an average of $11.9\%$ and achieves $22.6\%$ performance gain at the end of the simulation under incremental demand disruptions, as well as a gain of $11.5\%$ compared to MPC. The fact that the TTS difference curve is largely convex for MPC and concave for the antifragile RL algorithm can be quantified in Fig. \ref{fig: incr_d_sk_un}. While the final distribution skewness of the baseline and the antifragile RL algorithms is $0.73$ and $0.61$, respectively, the skewness of MPC records a value of $0.93$, indicating its high fragility under incremental demand disruptions. An interesting phenomenon worth noticing is the bump of the skewness curve of the proposed algorithm around episode $60-65$ in \ref{fig: incr_d_sk_un}, which can also be observed in Fig. \ref{fig: incr_d_sk}. The spike is likely due to the interplay between multiple RL hyperparameters. Particularly after episode $50$, the replay buffer contains a mixture of training data under both disrupted and undisrupted scenarios, resulting in learning from a mixed dataset and causing a moderate deviation from optimal performance.

Supply disruption can be modeled as a proportional reduction in both the capacity and the maximum vehicle accumulation on the MFD profile. Figure \ref{fig: s} presents the performance curves of the studied algorithms under such supply disruptions. In addition to MPC, baseline RL, and the proposed antifragile RL, the MPC-MHE variant in brown is also included. By modeling the supply disruption magnitude coefficient $\eta$ as an unknown system state variable, the MHE module estimates $\eta$ in real time to capture the extent of the MFD reduction. Fig. \ref{fig: incr_s} shows the performance curves of the four studied algorithms under no disruption for the first $50$ episodes, followed by linearly reduced MFD during episode $50-75$.

\afterpage{%
  \clearpage
    \begin{landscape} 
    \begin{figure}[htbp]
        \centering
        \subfigure[Performance curves in TTS.]{%
          {\includegraphics[width=0.485\textwidth]{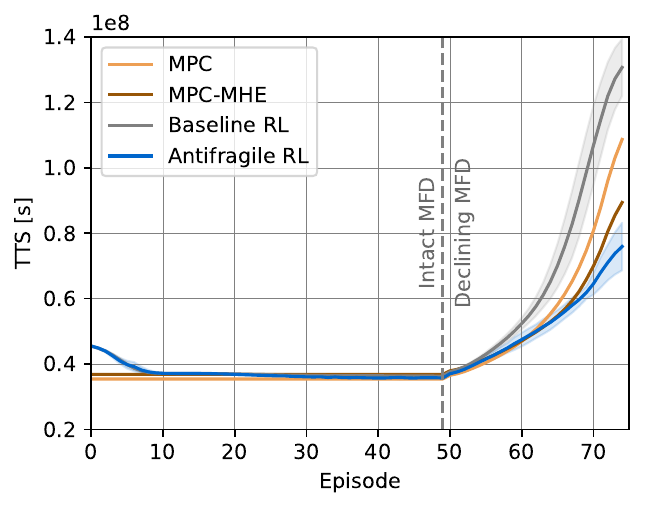}}
          \label{fig: incr_s}}
        \subfigure[Normalized over baseline RL.]{%
          {\includegraphics[width=0.43\textwidth]{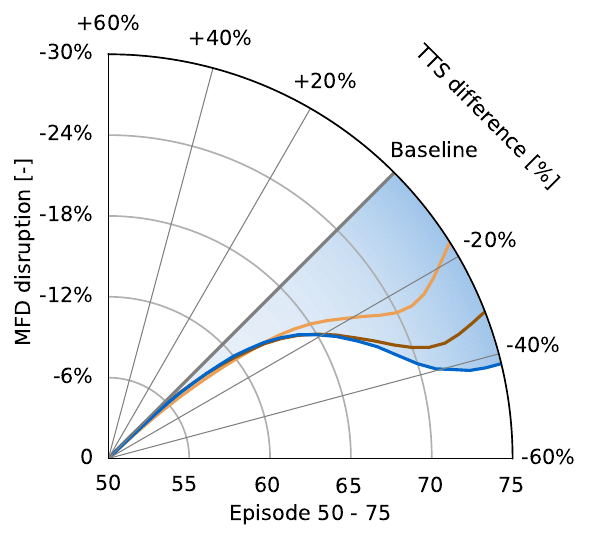}}
          \label{fig: incr_s_p}}
        \subfigure[Skewness up to the $n$-th episode.]{%
          {\includegraphics[width=0.369\textwidth]{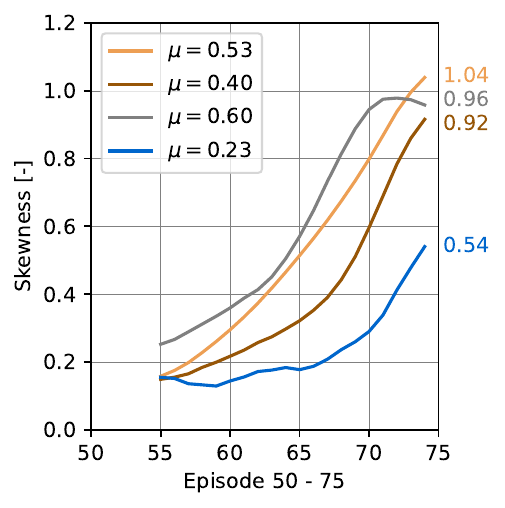}}
          \label{fig: incr_s_sk}}
        \caption{Performance curves under incremental supply disruptions.}
        \label{fig: s}
    
        \centering
        \subfigure[Performance curves in TTS.]{%
          {\includegraphics[width=0.485\textwidth]{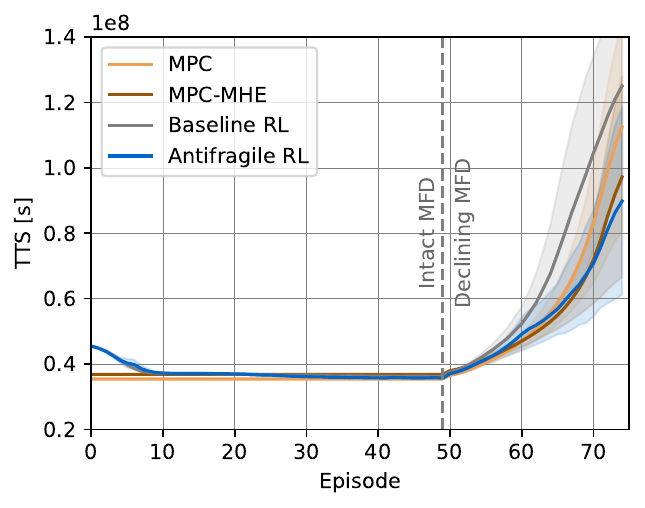}}
          \label{fig: incr_s_un}}
        \subfigure[Normalized over baseline RL.]{%
          {\includegraphics[width=0.43\textwidth]{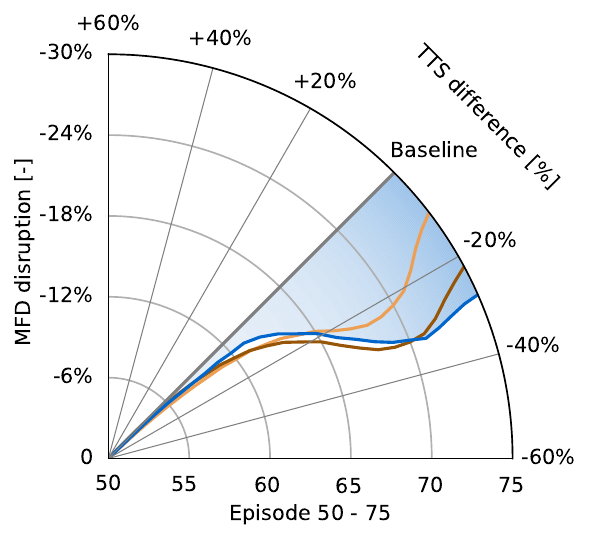}}
          \label{fig: incr_s_p_un}}
        \subfigure[Skewness up to the $n$-th episode.]{%
          {\includegraphics[width=0.369\textwidth]{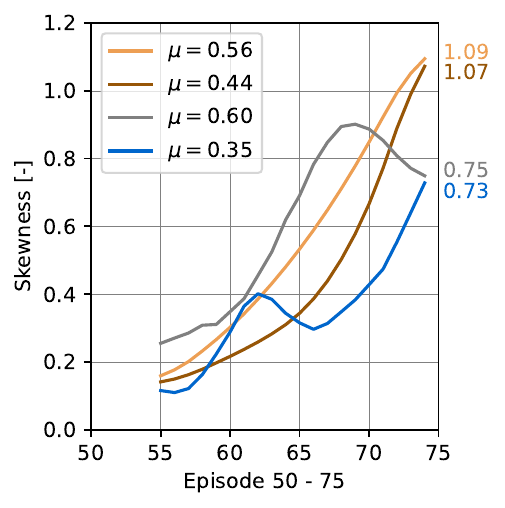}}
          \label{fig: incr_s_sk_un}}
        \caption{Performance curves under incremental supply disruptions with uncertainty.}
        \label{fig: s_un}
    \end{figure}
    \end{landscape}
  \clearpage
}

The performance gain is quantitatively analyzed in Fig. \ref{fig: incr_s_p}, which shows an average TTS reduction of $18.5\%$ from the proposed antifragile RL compared to the baseline RL algorithm, and reaches an ultimate performance improvement of $41.9\%$. It is worth noting that the MPC-MHE curve in brown also delivers strong performance, achieving a $31.6\%$ TTS reduction at peak disruption, highlighting the effectiveness of incorporating the supply disruption magnitude coefficient $\eta$ to be estimated within the MHE framework. Still, MHE relies on accurately modeling both the system dynamics and the evolution of the model reduction, which is highly challenging in real-world scenarios. 

The distribution skewness up to the $n$-th episode under disruption is demonstrated in Fig. \ref{fig: incr_s_sk}. The proposed antifragile RL method exhibits noticeably lower skewness across episodes $50-75$ compared to the other approaches, with a final skewness of $0.54$, whereas the skewness of the other three methods approaches or exceeds $1.0$. Note that the baseline RL algorithm exhibits a plateau in skewness toward the end of the simulation, which is caused by the fact that the network has already succumbed to full gridlock. And once the network is gridlocked, TTS merely increases linearly rather than exponentially with the rising vehicle accumulation. As a result, after surpassing a certain disruption threshold, the skewness of TTS initially plateaus and then gradually decreases toward $0$. Such a phenomenon is more pronounced in the following Fig. \ref{fig: incr_s_sk_un}, indicating that the proposed algorithm is only applicable under typical disruptions, while not suitable for handling catastrophic scenarios where the network rapidly descends into gridlock.

When supply disruption uncertainty is introduced into the simulation, as shown in Fig. \ref{fig: s_un}, the general trend is similar to that observed under demand disruption with uncertainty. The performance curves of MPC, MPC-MHE, and the antifragile RL algorithm all deteriorate to some extent in the presence of stochasticity. Interestingly, the performance of the baseline RL algorithm appears to improve at peak disruption compared to Fig. \ref{fig: incr_s}. Such improvement can be attributed to the onset of gridlock caused by major supply disruptions as well. When stochasticity amplifies a major supply disruption, its impact is marginal since the network is already gridlocked. On the contrary, when stochasticity leads to an attenuated coefficient, it reduces TTS and enhances performance. 

The performance gain of the proposed algorithm is demonstrated quantitatively in Fig. \ref{fig: incr_s_p_un}, showing an average improvement of $15.3\%$ and reaching $28.1\%$ at the final episode, while MPC and MPC-MHE show an ultimate gain of $10.0\%$ and $22.2\%$. Fig. \ref{fig: incr_s_sk_un} summarizes the skewness of the studied algorithms, with our proposed antifragile RL achieving the lowest value of $0.73$. Although the ultimate distribution skewness of the baseline RL reaches $0.75$, its peak skewness is $0.96$ at episode $70$, whereas the skewness of MPC and MPC-MHE both exceeds $1.0$.

\subsection{Performance under real-world limited observability}
\label{sec: limited_obs}

As obtaining real-world measurements of $n_{ij}$ and $q_{ij}$ is highly challenging, based on Eq. \ref{eq: full} and \ref{eq: limited} as the definition of states, we evaluate the performance of the baseline and proposed antifragile RL-based algorithms under conditions of real-world limited observability. It is noteworthy to highlight that the performance of the proposed algorithm under limited observability is not directly comparable to MPC or MPC-MHE under full observability, given their fundamentally distinct observability requirements. Fig. \ref{fig: d_obs} and \ref{fig: d_obs_un} illustrate the performance of algorithms in the presence of linearly increasing demand and supply disruptions, respectively.

\afterpage{%
  \clearpage
    \begin{landscape} 
    \begin{figure}[htbp]
    \centering
    \subfigure[Performance curves in TTS.]{%
      {\includegraphics[width=0.485\textwidth]{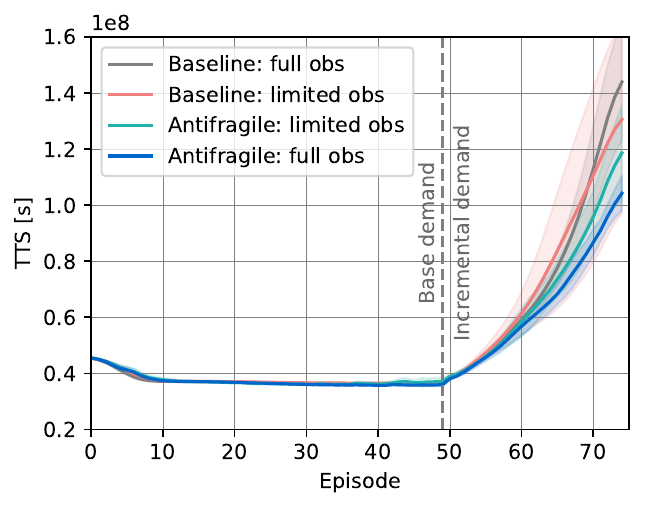}}
      \label{fig: incr_d_obs}}
    \subfigure[Normalized over baseline RL.]{%
      {\includegraphics[width=0.43\textwidth]{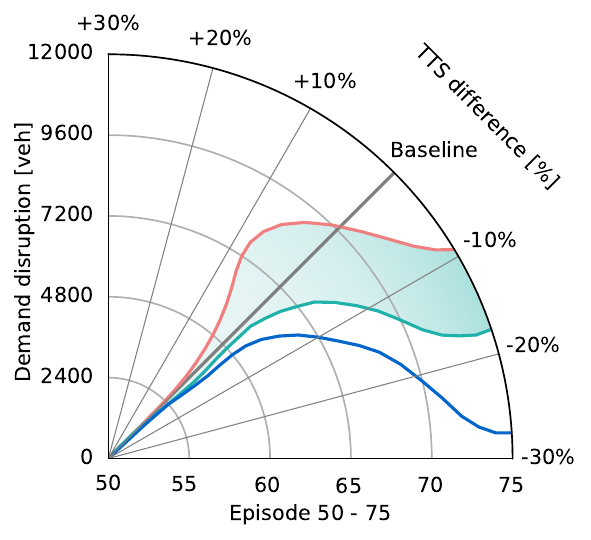}}
      \label{fig: incr_d_p_obs}}
    \subfigure[Skewness up to the $n$-th episode.]{%
      {\includegraphics[width=0.369\textwidth]{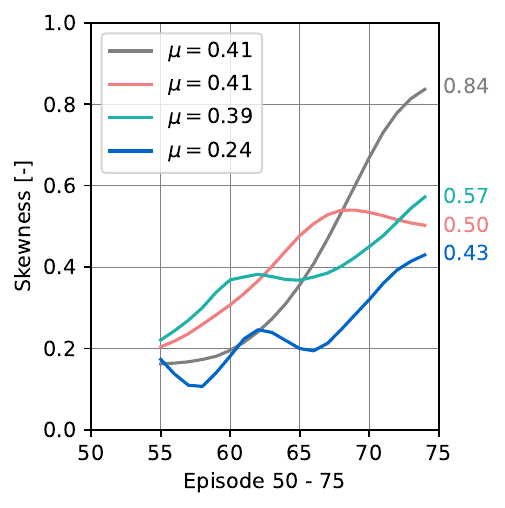}}
      \label{fig: incr_d_sk_obs}}
    \caption{Performance curves under incremental demand disruptions.}
    \label{fig: d_obs}

    \centering
    \subfigure[Performance curves in TTS.]{%
      {\includegraphics[width=0.485\textwidth]{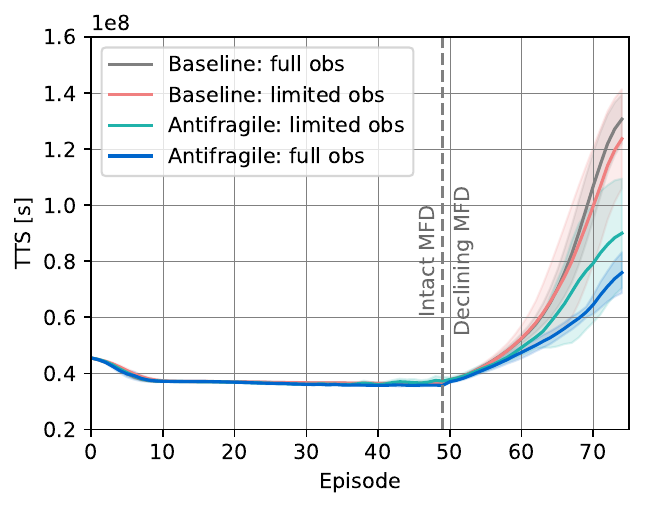}}
      \label{fig: incr_s_obs}}
    \subfigure[Normalized over baseline RL.]{%
      {\includegraphics[width=0.43\textwidth]{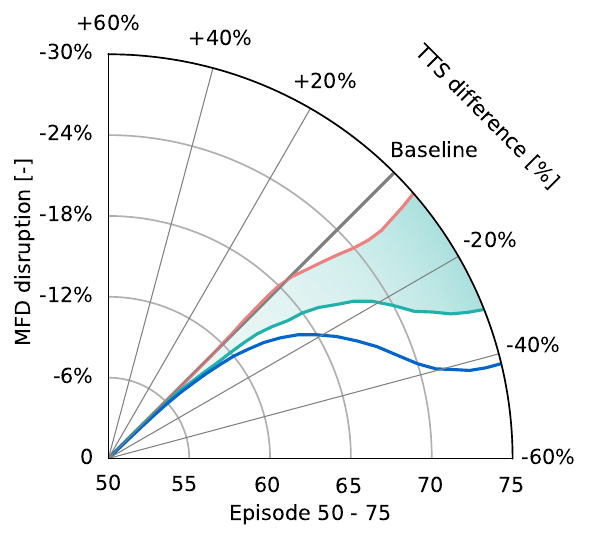}}
      \label{fig: incr_s_p_obs}}
    \subfigure[Skewness up to the $n$-th episode.]{%
      {\includegraphics[width=0.369\textwidth]{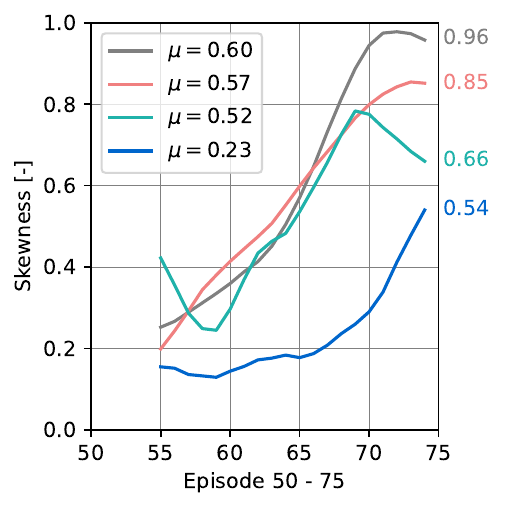}}
      \label{fig: incr_s_sk_obs}}
    \caption{Performance curves under incremental supply disruptions.}
    \label{fig: s_obs}
    \label{fig: d_obs_un}
    \end{figure}
    \end{landscape}
  \clearpage
}

In Fig. \ref{fig: incr_d_obs}, the performance curve of the antifragile RL algorithm under limited observability in light teal falls between that of the baseline RL and the antifragile RL with full observability, indicating that the algorithm sacrifices some performance in exchange for relaxing observability requirements. The TTS difference in Fig. \ref{fig: incr_d_p_obs} marks the performance difference between the baseline and antifragile RL-based algorithms under limited observability, taking an average of $4.8\%$ performance improvement and reaching $8.4\%$ at the end of the simulation. It should be highlighted that the performance trade-off to achieve realistic observability, i.e., the performance difference between limited (light teal) and full observability (blue) of the proposed algorithm, is an average of $4.6\%$. The distribution skewness is quantitatively illustrated in Fig. \ref{fig: incr_d_sk_obs}. Although the ultimate skewness of the proposed algorithm under limited observability is relatively high at $0.57$ at the final episode $75$, its average skewness $\mu=0.39$ lies between the skewness observed with full observability with $0.24$ and that of the baseline RL under limited observability with $0.41$.

In Fig. \ref{fig: s_obs}, the performance of both algorithms under supply disruptions with limited observability follows similar trends to those under demand disruption. The shaded area in Fig. \ref{fig: incr_d_p_obs} exhibits the TTS difference between the two algorithms with limited observability, which yields an average performance gain of $9.9\%$ and achieves a final value of $25.8\%$. The observability trade-off is $6.8\%$. Likewise, its average skewness $\mu=0.52$ between episodes $50$ and $75$ is also positioned between the baseline RL with limited observability $0.57$, and the antifragile RL with full observability $0.23$. 

\subsection{Explainability and sensitivity analysis}

Previous sections report performance across the full 75 training episodes for both the state-of-the-art and the proposed antifragile algorithms. Now we examine individual representative episodes to address the explainability of the two RL-based algorithms. Specifically, we present the perimeter control variables $u_{ij}$ from the RL agents and the consequent vehicle accumulation $n_i$. 

\begin{figure}[hbt!]
  \centering
    \subfigure[Episode 50 no disruption.]{%
      \includegraphics[width=0.36\textwidth]{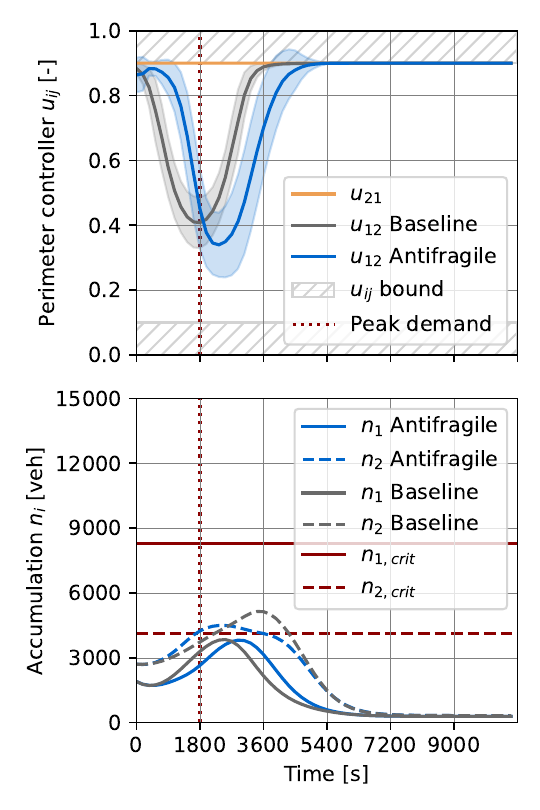}
      \label{fig: demand_50}}
    \subfigure[Episode 75 demand disruption.]{%
      \includegraphics[width=0.30\textwidth]{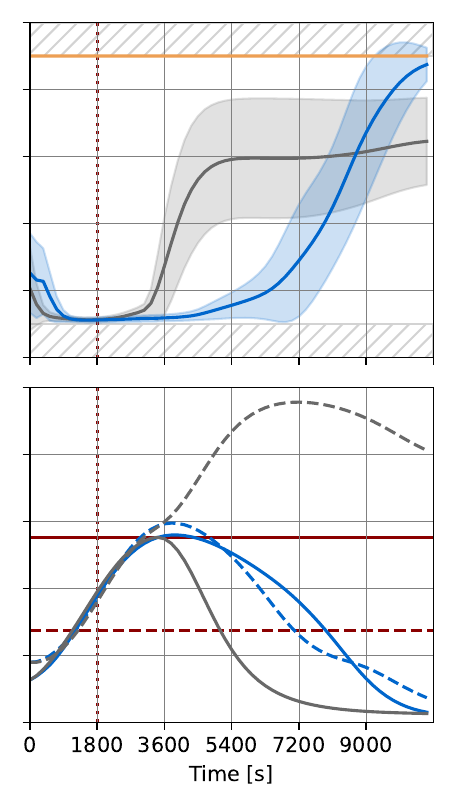}
      \label{fig: demand_75}}
    \subfigure[Episode 75 supply disruption.]{%
      \includegraphics[width=0.30\textwidth]{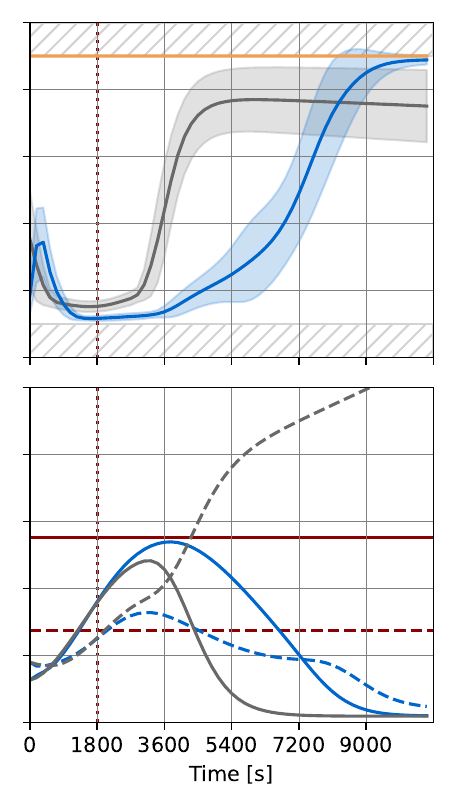}
      \label{fig: supply_75}}
    \caption{Representative episodes for the perimeter control variables and the vehicle accumulation under the last episode of no disruption, peak demand disruption, and peak supply disruption scenarios.}
    \label{fig: explainability}
\end{figure}

Fig. \ref{fig: explainability} highlights episodes 50 and 75, which are the final episode before any disruption is introduced and the final episode with the highest disruption magnitude. The upper plot in Fig. \ref{fig: demand_50} shows $u_{12}$ of both the baseline RL agent in gray and the proposed algorithm in blue, with the shaded area representing one standard deviation. Note that since the outer region is much less susceptible to traffic, both agents learn $u_{21}$  fairly easily to reach the upper bound $0.9$. The peak demand occurs at $t=1800$, resulting in the lowest $u_{12}$ for the baseline algorithm in gray. However, $n_{2}$ from the baseline algorithm depicted as the dashed gray curve in the lower plot has clearly exceeded the critical accumulation $n_{2, \mathrm{crit}}$ shown as the red dashed line, indicating the sub-optimal decision-making of $u_{12}$. Clearly, the learned policy of the baseline algorithm with state defined as $s_k=[n_{ij,k}, q_{ij,k}]$ is heavily influenced by $q_{ij,k}$, as $u_{12}$ from the baseline is nearly symmetric to the peak demand as in Fig. \ref{fig: inflow}. It should also be noted that, since MFD is largely a concave function, a small deviation between the proposed $n_2$ as the blue dashed curve and $n_{2, \mathrm{crit}}$ will only lead to minor performance loss, whereas an increasing deviation causes exponential loss. 

Fig. \ref{fig: demand_75} demonstrates $u_{ij}$ and $n_i$ of the final episode $75$ as the demand disruption magnitude gradually intensifies over episode $50-75$. The solid and dashed horizontal red lines represent the critical accumulation for either region $n_{1, \mathrm{crit}}$ or $n_{2, \mathrm{crit}}$, respectively. Due to the demand disruption on top of the base demand, both algorithms reach the lower bound $0.1$ before $t=1800$. The proposed RL agent remains approximately at the lower bound for a longer period, while the baseline RL agent quickly rebounds as a reaction to the waning of peak demand after $t=1800$, resulting in exceedingly high $n_2$ represented by the dashed gray curve. In contrast to Fig. \ref{fig: demand_75}, Fig. \ref{fig: supply_75} shows that the baseline $u_{12}$ in gray has not even reached the lower bound and rebounds quickly shortly after the peak of demand $t=1800$, while $u_{12}$ in blue remains at the lower bound first and only rises as the inner-region accumulation approaches the compromised critical accumulation due to the supply disruption. The above explainability analysis of the RL agent demonstrates that the proposed algorithm is capable of rejecting misinformation and thus adapting more effectively to disruptions than the baseline agent, leading to antifragile performance as the disruptions intensify.

As the major contribution of the present research is the design of the antifragile perimeter control framework, most hyperparameters of the proposed algorithm should largely align with the baseline in \cite{zhou_model-free_2021}, which achieves satisfactory convergence over training episodes. We further implement a sensitivity analysis on the reward weights, which is essential as a methodological contribution, and learning rates, as they are reasonably larger than in the original work, to maintain the learning capabilities under disruptions. Table \ref{tab: sensitivity} summarizes the TTS at the peak disruption magnitude and the maximal skewness. We tested different magnitudes of the weight parameters in the redundancy term and found $\omega_h=0.01$ and $\omega_{\Delta h}=0.02$ yield the best results with both the lowest TTS and skewness. As reward shaping heavily relies on a harmony of various terms in the reward function, the RL agent's decision-making can also be compromised when the weight parameters are too high. Note that the sensitivity analysis on the weight parameters is implemented under supply disruption scenarios to showcase its superior effectiveness. Results in Table \ref{tab: sensitivity} also demonstrate that $lr_a=0.008$ and $lr_c=0.004$ offer the most satisfactory overall performance under demand disruptions. Smaller learning rates will diminish the learning capability, especially under increasing disruptions. When the learning rates are high, although the performance and skewness can be even superior to the optimal learning rate, there is a higher likelihood of causing gradient explosion, which prevents convergence. 

\begin{table}[h]
\footnotesize
\centering
\renewcommand{\arraystretch}{1.5}
\caption{Sensitivity analysis on the redundancy weight parameters and the initial learning rates.}
\label{tab: sensitivity}
\setlength\tabcolsep{3pt} 
\begin{tabular}{C{4cm}C{2cm}C{1.6cm}|C{4cm}C{2.2cm}C{2cm}}
\toprule
\textbf{Weight parameters} & Performance & skewness & \textbf{Learning rate} & Performance & skewness 
\\ \midrule
$\omega_h=0.000, \omega_{\Delta h}=0.000$  & $7.84\cdot10^7$ & $0.67$  & $lr_a=0.002,lr_c=0.001$ & $2.233\cdot10^8$ & $0.98$
\\
\underline{$\omega_h=0.010, \omega_{\Delta h}=0.020$}  & $7.59\cdot10^7$ & $0.54$  & \underline{$lr_a=0.008,lr_c=0.004$} & $1.043\cdot10^8$ & $0.43$
\\
$\omega_h=1.000, \omega_{\Delta h}=2.000$  & $9.57\cdot10^7$ & $0.77$  & $lr_a=0.032,lr_c=0.016$ & $0.997\cdot10^8$/\texttt{NaN} & $0.40$/\texttt{NaN}
\\\bottomrule
\end{tabular}
\end{table}

Perimeter control is based on the MFD and is therefore intrinsically computationally efficient. Note that a full RL simulation episode consists of the scenario generation, training, and testing phases. Following the Ape-X architecture utilized in \citep{zhou_model-free_2021} and detailed in Section \ref{sec: ddpg}, the primary computational bottleneck resides in the scenario generation phase rather than training the neural network, where GPU acceleration offers marginal benefits. Implementation of the proposed algorithm on an Intel Core i7-1165G7 CPU (4 cores, 2.80 GHz), scenario generation requires an average of $55.4$ seconds, whereas training and testing phases are significantly faster at $5.5$ and $6.8$ seconds, respectively. In comparison, MPC and MPC-MHE take $5.2$ s and $9.7$ seconds to compute a single episode. Since the proposed antifragile RL algorithm involves a higher state dimension to accommodate derivative calculations, we further investigated the difference in the computational cost, which shows a moderate $18.7\%$ increase. Given the simple two-layer neural network architecture of the baseline, the total training time remains within seconds, rendering the increase negligible in practice.

\subsection{Antifragile module on recent RL algorithms}
So far, this study builds upon the DDPG agent and develops an antifragile perimeter control algorithm by augmenting the state space and reward shaping. The methodology functions as an add-on to the existing DDPG algorithm, enhancing its performance when faced with disruptions of previously unseen magnitudes. Meanwhile, however, recent advances in RL offer advantages including improved sample efficiency, training stability, and hyperparameter robustness, among others. Therefore, we apply the proposed methodology to state-of-the-art algorithms to assess whether, and to what extent, such algorithms would benefit from the inclusion of derivatives and reward shaping under increasingly severe disruptive events. In addition to DDPG, we consider Twin Delayed Deep Deterministic policy gradient (TD3), Soft Actor-Critic (SAC), and Proximal Policy Optimization (PPO) as baseline algorithms. TD3 employs a twin critics mechanism and takes the minimum Q-value estimate to mitigate overestimation, incorporates target policy smoothing to avoid chasing sharp peaks of Q-value, and delays actor updates so that the policy is trained on steadier value estimates \citep{fujimoto_addressing_2018}. SAC is a stochastic, off-policy actor–critic algorithm that maximizes the expected return augmented with an entropy bonus and applies twin Q-networks with automatic temperature tuning to achieve stable and sample-efficient exploration \citep{haarnoja_soft_2018}. PPO is an on-policy actor–critic algorithm that optimizes a clipped surrogate objective to prevent large policy updates, approximating trust region optimization without second-order derivatives while using generalized advantage estimation for stable, low-variance training \citep{schulman_proximal_2017}. All TD3, SAC, and PPO have seen growing applications in the discipline of reliability and risk engineering in recent years, such as in \cite{ding_soft_2025, cao_deep_2025, zhou_integrated_2026, walia2026uncertainty}.

\begin{figure}[hbt!]
  \centering
    \subfigure[Performance curves in TTS.]{%
      \includegraphics[width=0.43\textwidth]{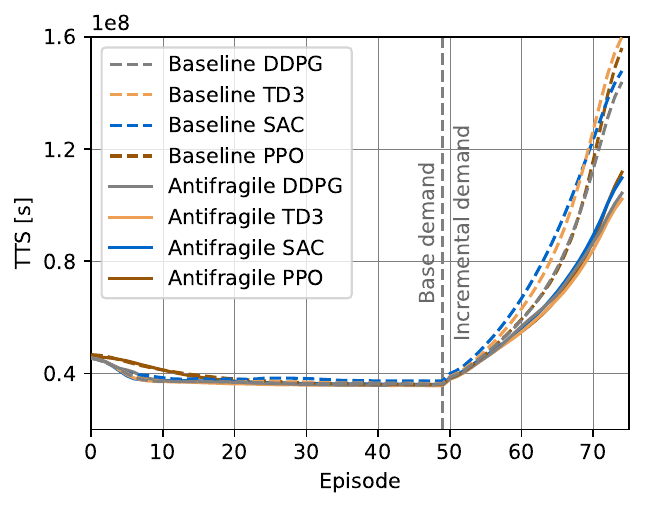}
      \label{fig: incr_d_sota}}
    \\
    \subfigure[Normalized over baseline DDPG.]{%
      \includegraphics[width=0.39\textwidth]{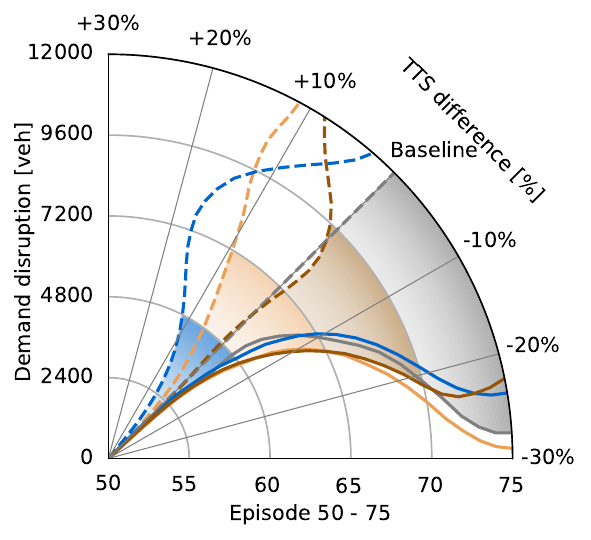}
      \label{fig: incr_d_p_sota}}
    \subfigure[Skewness up to the $n$-th episode.]{%
      \includegraphics[width=0.32\textwidth]{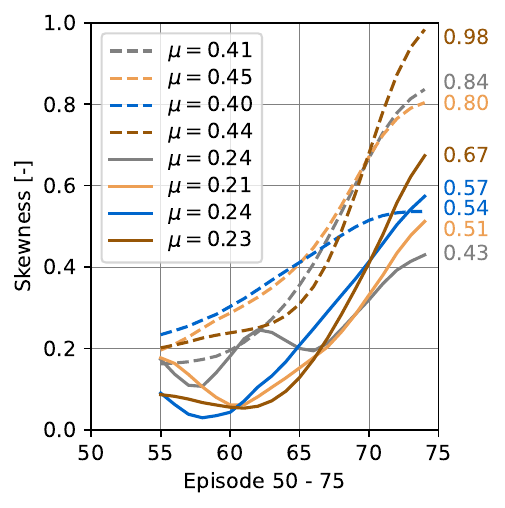}
      \label{fig: incr_d_sk_sota}}
    \caption{Performance curves and skewness under real-world disruptions.}
    \label{fig: sota}
\end{figure}

Fig. \ref{fig: sota} displays the performance curve in TTS and the distribution skewness of the investigated algorithms. The dashed curves mark the baseline algorithms for DDPG, TD3, SAC and PPO with $s_k=[n_{ij,k}, q_{ij,k}]$, while the solid curves represent the proposed antifragile variants with $s_k=[n_{ij, k}, \Delta n_{ij, k}, \Delta^2 n_{ij, k}, M_{ij, k}]$ and additional reward shaping terms defined as in Eq. \ref{eq: J_RL}. Clearly in Fig. \ref{fig: incr_d_sota}, the six performance curves under incremental demand disruptions between episode $50-75$ fall into two regimes, dependent on whether the proposed antifragile module has been leveraged. Fig. \ref{fig: incr_d_p_sota} further gauges the performance improvement between the baselines and their antifragile counterparts, showing a performance gain varying approximately between $25\% -35\%$. The distribution skewness is also illustrated in Fig. \ref{fig: incr_d_sk_sota}. The skewness of the solid curves is generally lower than that of the dashed curves. The average skewness $\mu$ of algorithms with antifragile modules is between $0.20-0.25$, compared to between $0.40-0.45$ for the baselines. The differences in the skewness demonstrate the antifragile properties of the proposed algorithm under disruptions of increasing magnitudes. Both the superior performance and the lower skewness have showcased the modularity and transferability of the proposed antifragile methodology regardless of the specific RL algorithm employed.

\subsection{Simulation study based on real-world data}
\label{sec: real-world}

\begin{figure}[hbt!]
  \centering
    \subfigure[The city center of Zurich.]{%
      \includegraphics[width=0.51\textwidth]{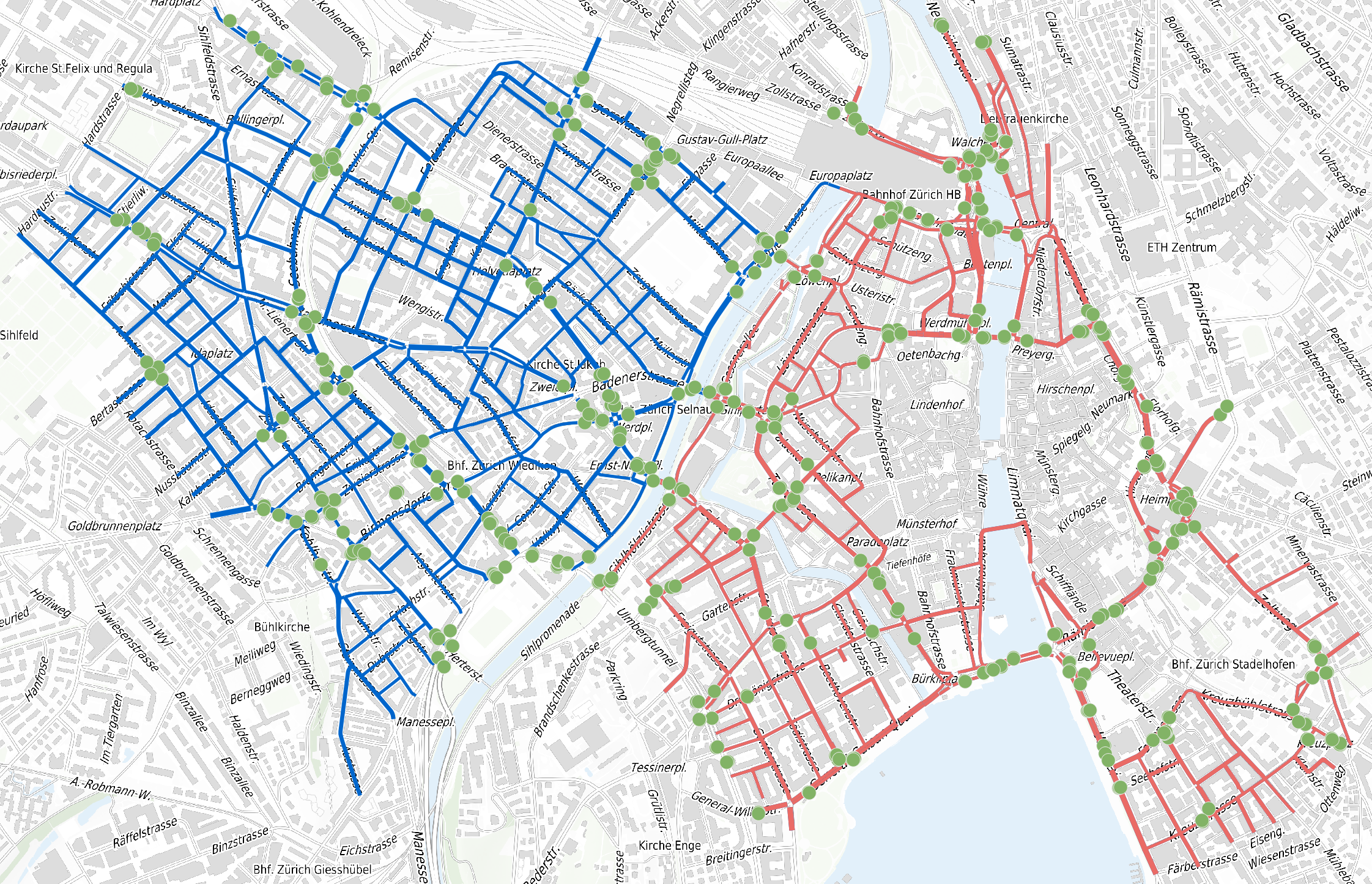}
      \label{fig: real_network}}
    \subfigure[Approximated MFDs with loop detector data.]{%
      \includegraphics[width=0.45\textwidth, trim=0 10 0 0]{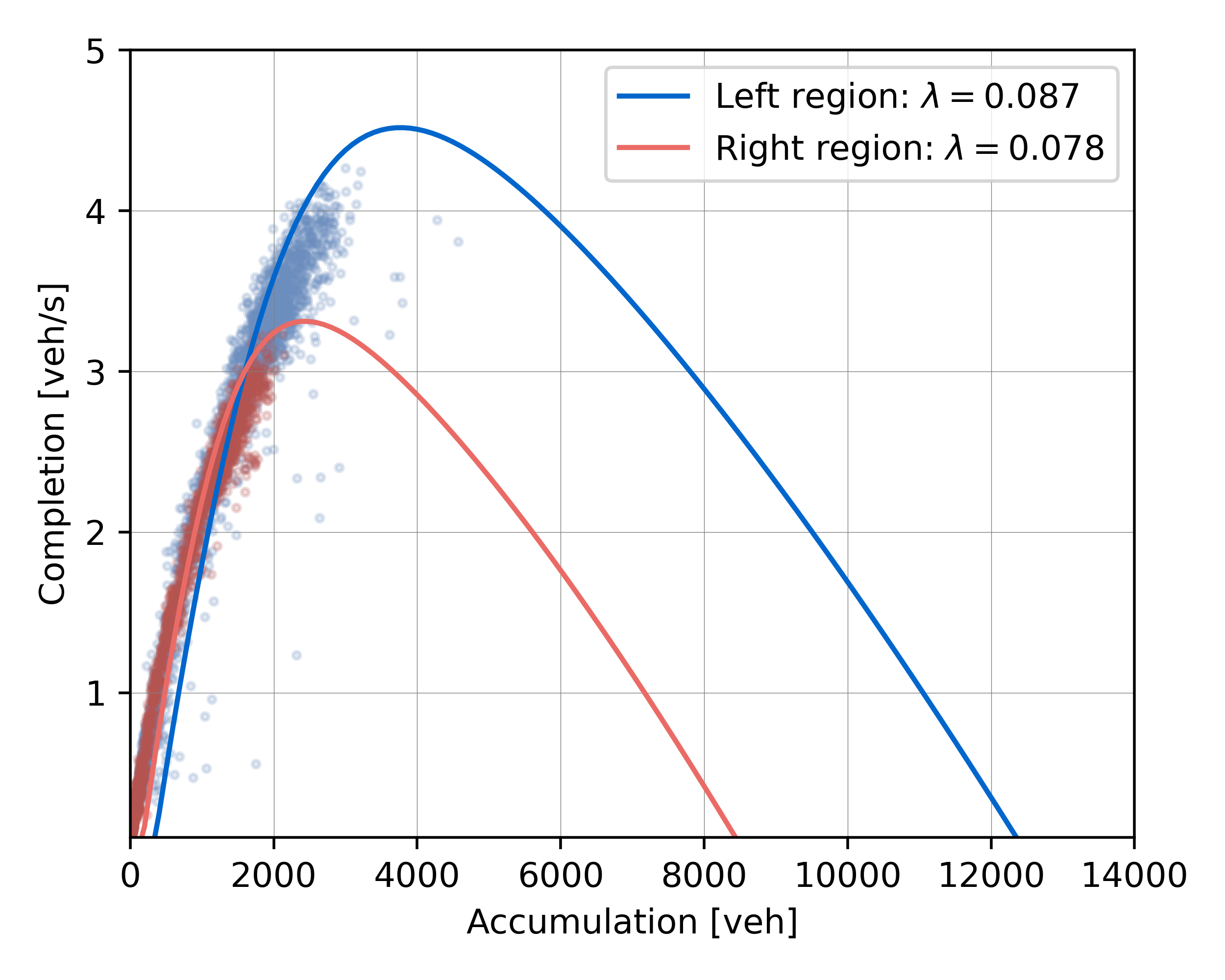}
      \label{fig: real_mfd}}
    \caption{Simulation environment constructed from the real-world network and measurements.}
    \label{fig: real_simulation}
\end{figure}

To demonstrate the applicability of the proposed algorithm in the real world and validate its robustness, especially under a time-varying demand profile, we further conduct a numerical simulation using data and models derived from real-world measurements. Instead of a cordon-shaped network, we utilize the network in \cite{fulton_impact_2025}, where the city center of Zurich is divided into left and right regions connected via bridges as bottlenecks. The two regions are represented by blue and red links for motorized vehicles in Fig. \ref{fig: real_network}. Such an experimental setup avoids the complexity of through traffic passing by the city center in real-world scenarios. The green dots represent the coordinates of 500 loop detectors spread across the studied region. The measurements from the relevant detectors are used to approximate the MFD for each region. Note that due to the insignificant congestion branch in Zurich, we substituted the third-degree polynomial approximation originally applied in \cite{zhou_model-free_2021} with the infrastructural potential $\lambda$ method introduced in \citep{ambuhl_functional_2020}, which is a semi-analytical method that guarantees the existence of a second root, as illustrated in Fig. \ref{fig: real_mfd}. The switch of the approximation function can also be used to demonstrate the robustness of our proposed algorithm to different MFD functional forms. The $\lambda$ method requires parameters with physical meanings, including free-flow speed, maximal capacity, backward wave speed, and maximal density, which are also provided in \citep{ambuhl_functional_2020} for the studied region. As described in \cite{sun_fragile_2024}, by taking the average trip length and total lane length into consideration, the flow-density MFD can be converted to the completion-accumulation MFD. 

\begin{figure}[hbt!]
  \centering
    \subfigure[Performance curves in TTS.]{%
      \includegraphics[width=0.55\textwidth]{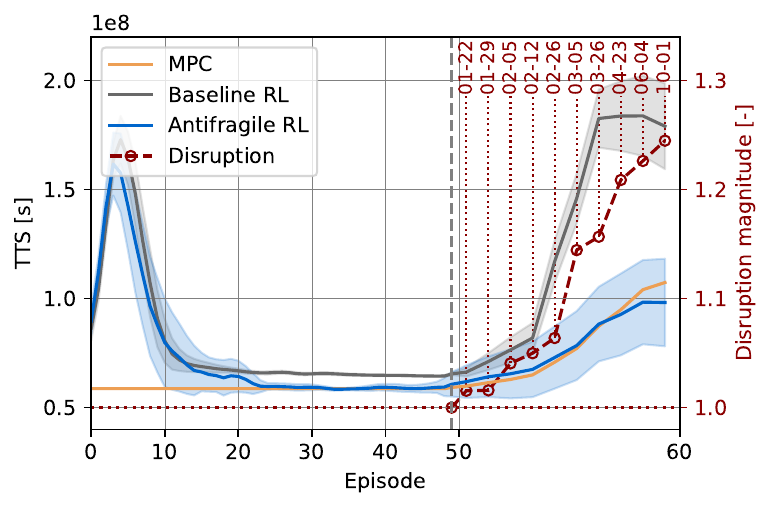}
      \label{fig: incr_d_real}}
    \subfigure[Skewness up to the $n$-th episode.]{%
      \includegraphics[width=0.405\textwidth]{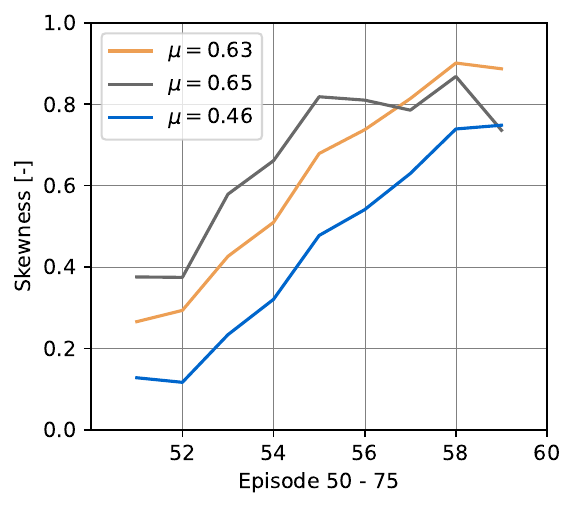}
      \label{fig: incr_d_sk_real}}
    \caption{Performance and skewness of the algorithms in the real-world network and measurements.}
    \label{fig: incr_real}
\end{figure}

A realistic morning peak hour demand profile in terms of OD pairs $q_{ij,k}$ is also provided in \cite{fulton_impact_2025} for the said network, which contains strong time-varying stochasticity. It should be noted that real-world OD data is near-impossible to acquire, and the demand profile is calibrated through the methodology described in \cite{ni2025estimation}. Due to the low level of inter-regional traffic flow, to ensure that perimeter control takes effect, a portion of intra-regional traffic within the right region is relocated with its new origin from the left region and destination to the right region. As both the supply and the demand based on real-world datasets in the city center of Zurich have been constructed, to showcase the antifragile properties of the proposed algorithm, realistic disruptions need to be approximated through historical data as well. First, the annual average traffic from the relevant loop detector measurements in 2020 is computed as the starting threshold and used as the initial training without disruptions. Then we search for the traffic flow from the beginning of 2021 that exceeds the previous threshold and update the threshold with the new peak. After searching for the whole year, a list of 10 events is summarized and illustrated by the dashed dark red curve with nodes in Fig. \ref{fig: incr_d_real}, with the maximal disruption magnitude as $24.5\%$ additional demand compared to the average traffic of the previous year. To tackle the time-varying demand dynamics, a moving window of 10 steps is leveraged to compute the first- and second-order differences in Eq. \ref{eq: full}. The performance curves from MPC, baseline RL, and the antifragile RL algorithm are presented in Fig. \ref{fig: incr_d_real}, where the proposed algorithm not only converges to a lower TTS compared to the baseline algorithm, but also yields an increasingly superior performance with growing disruptions calibrated with real-world measurements. Likewise, Fig. \ref{fig: incr_d_sk_real} illustrates the distribution skewness of the three examined algorithms over the last 10 episodes with disruptions. The proposed method achieves the lowest skewness overall, indicating its antifragile properties and the applicability under real-world scenarios.
   
\section{Conclusion}
\label{sec: conclusion}

As disruptions are ubiquitous in urban transportation systems and are expected to increase continuously with the ongoing urbanization, population growth, and more frequent natural hazards, this work introduces the novel concept of antifragility into traffic control to tackle the growing trend of disruptions and the fragile nature of road transportation networks. First, from a terminology perspective, antifragility is compared with other related concepts commonly applied in transportation, including robustness, resilience, reliability, and adaptiveness. Through reviewing previous research on applying RL algorithms to achieve robust, resilient, or reliable systems from either the reliability engineering or traffic engineering domains, we further explore how these properties can be induced with RL. Building on such concepts, we aim to induce antifragility in a state-of-the-art RL-based perimeter control algorithm by modifying the state definition and reward function. We incorporate the first- and second-order derivatives, representing the change rate and the curvature of traffic states, to provide the RL algorithm with richer and more reliable information under disruptive events. To mitigate the potential oscillations caused by integrating derivatives, a damping term is introduced to stabilize the computed actions. Furthermore, a redundant overcompensation term in the reward function strengthens the system's antifragility against disruptions. In addition to the performance gain and antifragility, another contribution of this work is the consideration of real-world observability constraints, complementing the drawback of the state-of-the-art algorithms in utilizing hardly observable sensor measurements.

We conducted comprehensive experiments to compare the proposed antifragile RL-based perimeter control algorithm against three baseline approaches: MPC, MPC-MHE, and a model-free DDPG algorithm. Two distinct manifestations of disruptions are examined, i.e., demand and supply disruptions. Uncertainties in disruption magnitude are also considered. The results first demonstrate the effectiveness of our proposed antifragile RL-based algorithm through its superior performance with lower TTS, ultimately achieving $27.6\%$ and $41.9\%$ performance improvements over the baseline RL algorithm under maximal demand and supply disruptions. More importantly, by adopting a novel method for quantifying antifragility through computing its performance distribution skewness, we confirm that the proposed algorithm exhibits greater antifragility with the lowest skewness among all the methods examined and delivers increasingly better performance as disruptions intensify. For example, the final skewness of the proposed antifragile RL algorithm under supply disruptions achieves $0.46$ compared to $1.04$ from MPC, $0.92$ from MPC-MHE, and $0.96$ from the baseline RL algorithm. Moreover, the performance and antifragile properties are studied under real-world limited observability, demonstrating that a certain degree of performance can be traded for using only accessible sensor measurements in reality. Moreover, since the proposed methodology functions as an add-on to the baseline DDPG algorithm, more advanced algorithms, including TD3 and SAC, are also investigated and compared. Results demonstrated the broad applicability of our proposed approach to other contemporary baselines. Lastly, we further validated the superior performance and relative antifragile properties of the proposed algorithm with a real-world case study, based on realistic demand, supply, and disruptions from the city center of Zurich.

Several limitations of the current study need to be addressed. For instance, this work primarily focuses on numerical simulations at the macroscopic level, with Section \ref{sec: real-world} showcasing a numerical simulation created from real-world data. It should be acknowledged that real-world operation with the proposed antifragile RL-based algorithm can be more accurately validated through traffic microsimulation software like SUMO. However, due to the complexity of selecting one or multiple functional strings of traffic lights to form the perimeter as well as dealing with potential spillover and gridlock at the perimeter gating, such works are generally carried out as independent research with additional techniques, such as minimizing relative queues in \cite{kampitakis_decentralized_2025}. Additionally, the current work considers only unimodal transportation as in \cite{ganin_resilience_2017}, and the algorithm can potentially be extended into an antifragile traffic control framework in a multimodal complex systems context \citep{li_adaptive_2025} for multi-reservoir transportation networks \citep{zhou_scalable_2023}. Moreover, measurement and model uncertainties are commonly studied when designing perimeter control algorithms to validate their robustness. However, to maintain a clear focus on antifragility rather than robustness, we confine our robustness evaluation within the investigation of the time-varying demand in the case study with real-world datasets. Sensor measurement errors, as in \citep{sirmatel_nonlinear_2020}, are not taken into account in this work. Nevertheless, with the proposed RL-based antifragile perimeter control framework, transportation networks can reach a trade-off between potential sensor error-induced minor performance degradation and major performance loss from disruptive events. Addressing these limitations is an essential step toward elevating the concept of antifragility and antifragile perimeter control from proof-of-concept to a full-fledged deployment-ready prototype. 

In conclusion, this study is the first of its kind to pioneer the application of antifragility in the operation and control of transportation systems, continuously improving the urban network performance under unforeseen disruptions with learning-based algorithms. Moreover, the concept is generic enough to be extended to other traffic control systems and potentially to control systems in general that are subject to growing disruptions.

\section*{CRediT authorship contribution statement}
Linghang Sun: Conceptualization, Investigation, Methodology, Visualization, Writing – original draft. Michail A. Makridis: Conceptualization, Methodology, Supervision, Writing - review \& editing. Alexander Genser: Methodology, Visualization, Writing - review \& editing. Cristian Axenie: Conceptualization, Project administration, Resources, Writing - review \& editing. Margherita Grossi: Project administration, Resources, Writing - review \& editing. Anastasios Kouvelas: Supervision, Writing - review \& editing.

\section*{Declaration of competing interest}
This research was kindly funded by the Huawei Munich Research Center under the framework of the Antigones project, with one of our co-authors being employed at the said company. Otherwise, the authors declare that they have no known competing financial interests or personal relationships that could have appeared to influence the work reported in this paper.

\section*{Acknowledgement}
This research was primarily funded by the Huawei Munich Research Center through the Antigones project. Additional financial support has been received from the Swiss State Secretariat for Education, Research, and Innovation (SERI) with contract number 25.00120, as part of the European Union's Horizon Europe project AntifragiCity with project ID 101203052.

\section*{Data availability}
The authors have no permission to share the data.

\bibliographystyle{elsarticle-harv} 
\bibliography{references}
\end{document}